\documentclass[journal,onecolumn]{IEEEtran}
\usepackage[nolist]{acronym} 
\usepackage{setspace}
\usepackage[final]{graphicx}
\usepackage{tikz}
\usepackage{pgfplots}
\usepgfplotslibrary{groupplots}
\usepackage{subfloat}
\usepackage{subfig}
\usepackage{caption}
\usepackage[dash,dot]{dashundergaps}
\usepackage{float}
\pgfplotsset{compat=newest}
\usepackage{units}
\usepackage{amsmath, amsbsy, amssymb}
\usepackage{tikzscale}
\usetikzlibrary{arrows}
\usepackage{color}
\usepackage{epigraph}
\usepackage{circuitikz}
\usepackage{multirow}
\usepackage[ruled,boxed,lined]{algorithm2e}
\usepackage[group-separator={,}]{siunitx}
\definecolor{darkgreen}{rgb}{0.125,0.5,0.169}
\definecolor{mittelblau}{RGB}{0, 126, 198}
\definecolor{violettblau}{cmyk}{0.9, 0.6, 0, 0}
\definecolor{rot}{RGB}{238, 28 35}
\definecolor{apfelgruen}{RGB}{140, 198, 62}
\definecolor{gelb}{RGB}{1, 221, 0}
\definecolor{orange}{RGB}{244, 111, 33}
\definecolor{pink}{RGB}{237, 0, 140}
\definecolor{lila}{RGB}{128, 10, 145}
\definecolor{hellgrau}{RGB}{224, 224, 224}
\definecolor{mittelgrau}{RGB}{128, 128, 128}
\definecolor{dunkelgrau}{RGB}{80,80,80}
\definecolor{anthrazit}{RGB}{19, 31, 31}

\pgfplotscreateplotcyclelist{corporate colours markers}{%
rot, every mark/.append style={fill=.!80!rot},mark=*\\%
mittelblau, every mark/.append style={fill=.!80!mittelblau},mark=square*\\%
apfelgruen, every mark/.append style={fill=.!80!apfelgruen},mark=triangle*\\%
orange, mark=star\\%
pink, every mark/.append style={fill=.!80!pink},mark=diamond*\\%
violettblau, every mark/.append style={fill=.!80!violettblau},mark=otimes*\\%
lila, mark=|\\%
gelb, every mark/.append style={fill=.!80!gelb},mark=pentagon*\\%
hellgrau, mark=text,text mark=p\\%
anthrazit, mark=text,text mark=a\\%
}

\pgfplotscreateplotcyclelist{corporate colours markers double}{%
rot, every mark/.append style={fill=.!80!rot},mark=*\\%
rot, dashed, every mark/.append style={fill=.!80!rot,solid},mark=*\\%
mittelblau, every mark/.append style={fill=.!80!mittelblau},mark=square*\\%
mittelblau, dashed, every mark/.append style={fill=.!80!mittelblau,solid},mark=square*\\%
apfelgruen, every mark/.append style={fill=.!80!apfelgruen},mark=triangle*\\%
apfelgruen, dashed, every mark/.append style={fill=.!80!apfelgruen,solid},mark=triangle*\\%
orange, mark=star\\%
orange, dashed, mark=star\\%
pink, every mark/.append style={fill=.!80!pink},mark=diamond*\\%
pink, dashed, every mark/.append style={fill=.!80!pink,solid},mark=diamond*\\%
violettblau, every mark/.append style={fill=.!80!violettblau},mark=otimes*\\%
violettblau, dashed, every mark/.append style={fill=.!80!violettblau,solid},mark=otimes*\\%
lila, mark=|\\%
lila, dashed, mark=|\\%
gelb, every mark/.append style={fill=.!80!gelb},mark=pentagon*\\%
gelb, dashed, every mark/.append style={fill=.!80!gelb,solid},mark=pentagon*\\%
}
\usetikzlibrary{shapes,arrows}
\begin{document}

\title{Enabling FDD Massive MIMO through Deep Learning-based Channel Prediction}

\author{Maximilian Arnold, Sebastian D\"orner, Sebastian Cammerer, Sarah Yan,\\ Jakob Hoydis, and Stephan ten Brink%
\thanks{M.~Arnold, S.~D\"orner, S.~Cammerer, S.~Yan, and S. ten Brink are with the Institute of Telecommunications, University of  Stuttgart, Pfaffenwaldring 47, 70659 Stuttgart, Germany, \{arnold,doerner,cammerer,tenbrink\}@inue.uni-stuttgart.de)}.%
\thanks{J.~Hoydis is with Nokia Bell Labs, Route de Villejust, 91620 Nozay, France, jakob.hoydis@nokia-bell-labs.com.}}

\maketitle
\renewcommand{\vec}[1]{\mathbf{#1}}
\newcommand{\vecs}[1]{\boldsymbol{#1}}

\newcommand{\av}{\vec{a}}
\newcommand{\bv}{\vec{b}}
\newcommand{\cv}{\vec{c}}
\newcommand{\dv}{\vec{d}}
\newcommand{\ev}{\vec{e}}
\newcommand{\fv}{\vec{f}}
\newcommand{\gv}{\vec{g}}
\newcommand{\hv}{\vec{h}}
\newcommand{\iv}{\vec{i}}
\newcommand{\jv}{\vec{j}}
\newcommand{\kv}{\vec{k}}
\newcommand{\lv}{\vec{l}}
\newcommand{\mv}{\vec{m}}
\newcommand{\nv}{\vec{n}}
\newcommand{\ov}{\vec{o}}
\newcommand{\pv}{\vec{p}}
\newcommand{\qv}{\vec{q}}
\newcommand{\rv}{\vec{r}}
\newcommand{\sv}{\vec{s}}
\newcommand{\tv}{\vec{t}}
\newcommand{\uv}{\vec{u}}
\newcommand{\vv}{\vec{v}}
\newcommand{\wv}{\vec{w}}
\newcommand{\xv}{\vec{x}}
\newcommand{\yv}{\vec{y}}
\newcommand{\zv}{\vec{z}}
\newcommand{\zerov}{\vec{0}}
\newcommand{\onev}{\vec{1}}
\newcommand{\alphav}{\vecs{\alpha}}
\newcommand{\betav}{\vecs{\beta}}
\newcommand{\gammav}{\vecs{\gamma}}
\newcommand{\lambdav}{\vecs{\lambda}}
\newcommand{\omegav}{\vecs{\omega}}
\newcommand{\sigmav}{\vecs{\sigma}}
\newcommand{\tauv}{\vecs{\tau}}

\newcommand{\Am}{\vec{A}}
\newcommand{\Bm}{\vec{B}}
\newcommand{\Cm}{\vec{C}}
\newcommand{\Dm}{\vec{D}}
\newcommand{\Em}{\vec{E}}
\newcommand{\Fm}{\vec{F}}
\newcommand{\Gm}{\vec{G}}
\newcommand{\Hm}{\vec{H}}
\newcommand{\Id}{\vec{I}}
\newcommand{\Jm}{\vec{J}}
\newcommand{\Km}{\vec{K}}
\newcommand{\Lm}{\vec{L}}
\newcommand{\Mm}{\vec{M}}
\newcommand{\Nm}{\vec{N}}
\newcommand{\Om}{\vec{O}}
\newcommand{\Pm}{\vec{P}}
\newcommand{\Qm}{\vec{Q}}
\newcommand{\Rm}{\vec{R}}
\newcommand{\Sm}{\vec{S}}
\newcommand{\Tm}{\vec{T}}
\newcommand{\Um}{\vec{U}}
\newcommand{\Vm}{\vec{V}}
\newcommand{\Wm}{\vec{W}}
\newcommand{\Xm}{\vec{X}}
\newcommand{\Ym}{\vec{Y}}
\newcommand{\Zm}{\vec{Z}}
\newcommand{\Lambdam}{\vecs{\Lambda}}
\newcommand{\Pim}{\vecs{\Pi}}

\newcommand{\Ac}{{\cal A}}
\newcommand{\Bc}{{\cal B}}
\newcommand{\Cc}{{\cal C}}
\newcommand{\Dc}{{\cal D}}
\newcommand{\Ec}{{\cal E}}
\newcommand{\Fc}{{\cal F}}
\newcommand{\Gc}{{\cal G}}
\newcommand{\Hc}{{\cal H}}
\newcommand{\Ic}{{\cal I}}
\newcommand{\Jc}{{\cal J}}
\newcommand{\Kc}{{\cal K}}
\newcommand{\Lc}{{\cal L}}
\newcommand{\Mc}{{\cal M}}
\newcommand{\Nc}{{\cal N}}
\newcommand{\Oc}{{\cal O}}
\newcommand{\Pc}{{\cal P}}
\newcommand{\Qc}{{\cal Q}}
\newcommand{\Rc}{{\cal R}}
\newcommand{\Sc}{{\cal S}}
\newcommand{\Tc}{{\cal T}}
\newcommand{\Uc}{{\cal U}}
\newcommand{\Wc}{{\cal W}}
\newcommand{\Vc}{{\cal V}}
\newcommand{\Xc}{{\cal X}}
\newcommand{\Yc}{{\cal Y}}
\newcommand{\Zc}{{\cal Z}}

\newcommand{\CN}{\Cc\Nc}

\newcommand{\CC}{\mathbb{C}}
\newcommand{\MM}{\mathbb{M}}
\newcommand{\NN}{\mathbb{N}}
\newcommand{\RR}{\mathbb{R}}

\newcommand{\htp}{^{\mathsf{H}}}
\newcommand{\tp}{^{\mathsf{T}}}

\newcommand{\LB}{\left(}
\newcommand{\RB}{\right)}
\newcommand{\LP}{\left\{}
\newcommand{\RP}{\right\}}
\newcommand{\LSB}{\left[}
\newcommand{\RSB}{\right]}

\renewcommand{\ln}[1]{\mathop{\mathrm{ln}}\LB #1\RB}
\newcommand\norm[1]{\left\lVert#1\right\rVert}
\newcommand{\cs}[1]{\mathop{\mathrm{cs}}\LSB #1\RSB}

\newcommand{\EE}{{\mathbb{E}}}
\newcommand{\Expect}[2]{\EE_{#1}\LSB #2\RSB}

\newtheorem{definition}{Definition}[section]
\newtheorem{remark}{Remark}

\begin{acronym}
 \acro{CSI}{channel state information}
 \acro{UE}{user equipment}
 \acro{UL}{uplink}
 \acro{BS}{basestation}
 \acro{TDD}{time division duplex}
 \acro{FDD}{frequency division duplex}
 \acro{ECC}{error-correcting code}
 \acro{MLD}{maximum likelihood decoding}
 \acro{HDD}{hard decision decoding}
 \acro{IF}{intermediate frequency}
 \acro{RF}{radio frequency}
 \acro{SDD}{soft decision decoding}
 \acro{NND}{neural network decoding}
 \acro{CNN}{convolutional neural network}
 \acro{ML}{maximum likelihood}
 \acro{GPU}{graphical processing unit}
 \acro{BP}{belief propagation}
 \acro{LTE}{Long Term Evolution}
 \acro{BER}{bit error rate}
 \acro{SNR}{signal-to-noise-ratio}
 \acro{ReLU}{rectified linear unit}
 \acro{BPSK}{binary phase shift keying}
 \acro{QPSK}{quadrature phase shift keying}
 \acro{AWGN}{additive white Gaussian noise}
 \acro{MSE}{mean squared error}
 \acro{LLR}{log-likelihood ratio}
 \acro{MAP}{maximum a posteriori}
 \acro{NVE}{normalized validation error}
 \acro{BCE}{binary cross-entropy}
 \acro{BLER}{block error rate}
 \acro{SQR}{signal-to-quantisation-noise-ratio}
 \acro{MIMO}{multiple-input multiple-output}
 \acro{OFDM}{orthogonal frequency division multiplex}
 \acro{RF}{radio frequency}
 \acro{LOS}{line of sight}
 \acro{NLoS}{non-line of sight}
 \acro{NMSE}{normalized mean squared error}
 \acro{CFO}{carrier frequency offset}
 \acro{SFO}{sampling frequency offset}
 \acro{IPS}{indoor positioning system}
 \acro{TRIPS}{time-reversal IPS}
 \acro{RSSI}{received signal strength indicator}
 \acro{MIMO}{multiple-input multiple-output}
 \acro{ENoB}{effective number of bits}
 \acro{AGC}{automated gain control}
 \acro{ADC}{analog to digital converter}
 \acro{ADCs}{analog to digital converters}
 \acro{FB}{front bandpass}
 \acro{FPGA}{field programmable gate array}
 \acro{JSDM}{Joint Spatial Division and Multiplexing}
 \acro{NN}{neural network}
 \acro{IF}{intermediate frequency}
 \acro{LoS}{line-of-sight}
 \acro{NLoS}{non-line-of-sight}
 \acro{DSP}{digital signal processing}
 \acro{AFE}{analog front end}
 \acro{SQNR}{signal-to-quantisation-noise-ratio}
 \acro{SINR}{signal-to-interference-noise-ratio}
 \acro{ENoB}{effective number of bits}
 \acro{AGC}{automated gain control}
 \acro{PCB}{printed circuit board}
 \acro{EVM}{error vector mangnitude}
 \acro{CDF}{cumulative distribution function}
 \acro{MRC}{maximum ratio combining}
 \acro{MRP}{maximum ratio precoding}
 \acro{MRT}{maximum ratio transmission}
 \acro{DeepL}{deep-learning}
 \acro{DL}{downlink}
 \acro{SISO}{single-input single-output}
 \acro{SGD}{stochastic gradient descent}
 \acro{CP}{cyclic prefix}
 \acro{MISO}{Multiple Input Single Output}
 \acro{LMMSE}{linear minimum mean square error}
 \acro{ZF}{zero forcing}
 \acro{USRP}{universal software radio peripheral}
\end{acronym}

\begin{abstract}
A major obstacle for widespread deployment of \ac{FDD}-based Massive \ac{MIMO} communications is the large signaling overhead for reporting full \ac{DL} \ac{CSI} back to the \ac{BS}, in order to enable closed-loop precoding.
We completely remove this overhead by a deep-learning based channel extrapolation (or ``prediction'') approach and demonstrate that a \ac{NN} at the \ac{BS} can infer the \ac{DL} \ac{CSI} centered around a frequency $f_{\text{DL}}$ by solely observing \ac{UL} \ac{CSI} on a different, yet adjacent frequency band around $f_{\text{UL}}$; no more pilot/reporting overhead is needed than with a genuine \ac{TDD}-based system.
The rationale is that scatterers and the large-scale propagation environment are sufficiently similar to allow a \ac{NN} to learn about the physical connections and constraints between two neighboring frequency bands, and thus provide a well-operating system even when classic extrapolation methods, like the Wiener filter (used as a baseline for comparison throughout) fails.
We study its performance for various state-of-the-art Massive \ac{MIMO} channel models, and, even more so, evaluate the scheme using actual Massive \ac{MIMO} channel measurements, rendering it to be practically feasible at negligible loss in spectral efficiency when compared to a genuine \ac{TDD}-based system.
\end{abstract}

\acresetall

\section{Introduction}

With a significant increase in area throughput, Massive \ac{MIMO} antenna communication has become an enabling technology for the upcoming fifth generation (5G) wireless mobile communication systems \cite{EmilBjoernson2017,HoydisTenBrinkMMIMO,MMIMO5GCommMag,5GTechCommMag}.
However, Massive \ac{MIMO} systems described in current research literature commonly exploit channel reciprocity and hence rely on \ac{TDD}-based approaches \cite{EmilBjoernson2017}, i.e., \ac{UL} and \ac{DL} channels share the \emph{same} frequency band in orthogonal time intervals.
Achieving such reciprocity in practice requires accurate hardware with costly calibration circuitry.
To mitigate this issue, various alternatives to a full Massive \ac{MIMO} system have been proposed such as the \emph{grid of beams} \cite{7881048} and codebook Massive \ac{MIMO} \cite{DLChannelFDD}. 
As \ac{FDD} does not rely on channel reciprocity, it is a practically highly relevant topic, and both academia and industry have been striving for enabling \ac{FDD} Massive MIMO.
However, it is widely accepted as a fundamental limitation that \ac{FDD} Massive \ac{MIMO} has a prohibitive high piloting/reporting overhead related to channel estimation in both \ac{UL} and \ac{DL} direction \cite{EmilBjoernson2017}, unless certain conditions on the channel covariance structure are satisfied which, however, are rarely observed in practice \cite{6310934,8094949}.
In \ac{TDD}, the \ac{UE} sends a single pilot symbol to its $M$ antennas at the \ac{BS}, allowing the \ac{BS} to estimate the \ac{UL} channels in one time slot, and, by exploiting channel reciprocity, to re-use these \ac{UL} channel estimates for precoding in the \ac{DL}, without the need for any costly \ac{CSI} reporting feedback.
However, in \ac{FDD} systems, this reciprocity does not hold from one frequency band to another and, thus, the \ac{DL} channels from each antenna to the \ac{UE} must be estimated separately, and reported back to the \ac{BS}, incurring an overhead proportional to the number of antennas $M$ \cite{EmilBjoernson2017}.

\begin{figure}[H]
	\centering
	\includegraphics{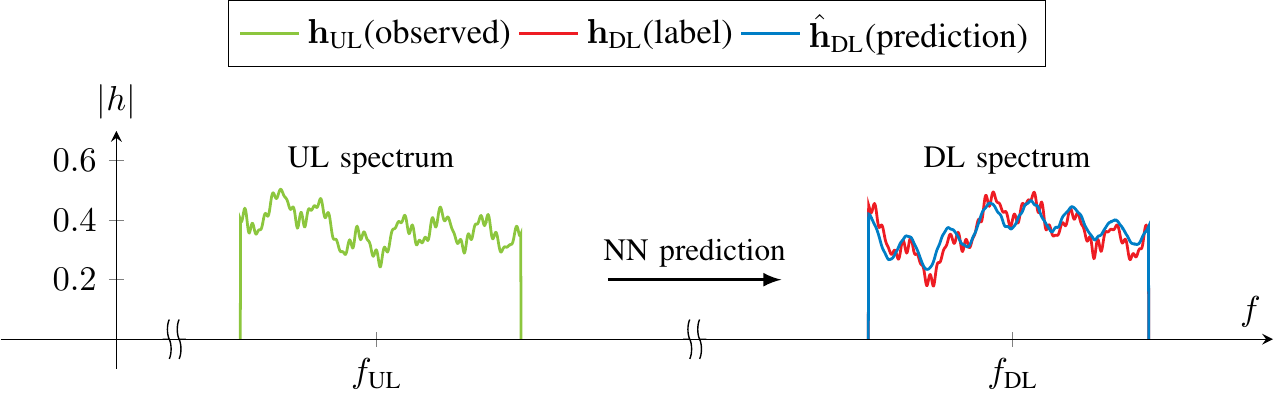}
	\caption{Using the uplink \ac{CSI} at $f_{\text{UL}}$ to infer on downlink \ac{CSI} at $f_{\text{DL}}$.} 
	\label{fig:Est-Princ}
\end{figure}

On the other hand, \ac{FDD} Massive \ac{MIMO} promises compatibility to current frequency allocations for mobile communication systems and, potentially, reduces latency \cite{7402270}.
It has been shown that a compressed sensing approach \cite{8284057} achieves almost the same minimal piloting overhead as \ac{TDD} Massive \ac{MIMO}.
However, it is unclear whether the general assumption of channel sparsity holds in practice \cite{6310934,7008286}.
Other approaches such as \cite{Vasisht2016,Yang2018} rely on an underlying channel model including expert knowledge on the antenna geometry.
Further, the authors in \cite{Soltani2018} propose a super-resolution channel estimation via deep learning in such a way that the \ac{NN} interpolates subcarriers without pilots from neighboring subcarriers.
In \cite{8052521} it was proposed to replace an \ac{OFDM} channel estimation combined with data detection by a \ac{NN}.
Recently, different \ac{CSI} prediction approaches exploiting channel sparsity have been proposed in \cite{ZhiyuanJiang2018,RuichenDeng2018}.
Another particular interesting idea to reduce the piloting overhead by using a subset of antennas to predict the remaining antennas was introduced in \cite{PeihaoDong2018}. 
While this approach aims at predicting the channel at a different spatial location, but on the same frequency, we try, in this work, to estimate the channel at a different frequency but for the same spatial position.
However, none of these works focus on the problem of extrapolating/prediction from one frequency band to an adjacent frequency band and, to the best of our knowledge, no general solution for the \ac{FDD} Massive \ac{MIMO} problem is known.
A \emph{classical} estimation using the channel covariance matrix, which has not resulted a practical solution yet, was approached in \cite{DLChannelFDD,DLChannelCovarianceFDDMMIMO}. 
Although machine learning and, in particular, deep learning has attracted a lot of attention for a wide variation of different communication applications, it currently lacks of solutions showing practical gains.
One important contribution of our work is to show that a \ac{NN} can \emph{enable} solutions (i.e., \ac{FDD} \ac{CSI} extrapolation/prediction) for which simply no satisfactory \emph{classical} signal processing algorithm is known, or, even more so, where classic approaches (Wiener filter) would fail.

In this work, the general idea is to estimate the \ac{DL} \ac{CSI} only based on the observed \ac{UL} \ac{CSI} via deep \acp{NN} and, thus, overcome a major practical limitation for \ac{FDD} Massive \ac{MIMO}.
Our approach works for any considered scenario with adjacent frequency bands and, in contrast to compressed sensing, does not rely on channel sparsity as it also works in the \ac{SISO} scenario. 
The intuition behind this prediction scheme is that the impinging waves at the \ac{UE}, generated through different scatterers, are assumed to be similar over a small frequency bandwidth, although it may be hard to describe the connection in terms of an analytical model. Therefore, a \ac{NN} should be able to \emph{approximate} such a function without the need of any a priori knowledge other than the observed measured data.
Fig.~\ref{fig:Est-Princ} illustrates the proposed scheme for an \ac{OFDM} system, extrapolating \ac{CSI} from a frequency band centered around $f_{\text{UL}}$ to a nearby frequency band centered around $f_{\text{DL}}$.  

This ease in modeling and flexibility in application comes at the cost of acquisitioning large quantities of channel estimates, plus retraining whenever the radio environment changes significantly.
We believe, however, that data acquisition is not a real limitation, as a hybrid approach could use pilots whenever the \ac{BS} operates at low average load without any additional effort, allowing the continuous refinement of the \ac{NN} weights; only during high load periods the \ac{BS} may switch to the proposed \ac{CSI} prediction scheme.
Also, the training could be outsourced to a datacenter and only the updated weights need to be transmitted to the \ac{BS} whenever a performance improvement can be achieved.

In a first step, we analyze the proposed scheme with the help of several simple examples, such as the \ac{SISO} setting to demonstrate the general feasibility of \ac{CSI} prediction from \ac{UL} to \ac{DL} in the \ac{SISO} scenario.
This provides important insights into designing and training a \ac{NN} for such a signal processing task, and also allows the comparison within an analytical setup.
Next, standardized channel models such as the 3GPP 38.901 channel model (implemented by the Quadriga framework \cite{Jaeckel2014}) and measurements are evaluated to show the viability of our proposed setup in a realistic setting.
Furthermore, with our measurements we show that the spectral efficiency loss in a practical system with 32 antennas is rather moderate $(\sim$10\% in \ac{LoS}, $\sim$20\% in \ac{NLoS}), rendering our scheme to be an attractive proposition for future wireless communication systems.

The remainder of this paper is structured as follows:
Section~\ref{sec:model_and_limits} starts with fundamental limitations of \ac{FDD} Massive \ac{MIMO} and discusses different metrics needed to evaluate predicted \ac{CSI}.
Section~\ref{sec:deep_learning} provides a short introduction to deep learning and clarifies notation.
In Section~\ref{sec:siso}, we introduce the proposed \ac{NN}-based approach for the \ac{SISO} scenario, provide results for simulated and measured scenarios and compare the \ac{NN}'s performance to a Wiener filter-based approach, i.e., \emph{the} classic signal processing baseline.
The simulations and measurements are then extended to a Massive \ac{MIMO} scenario in Section~\ref{sec:mimo}; finally Section~\ref{sec:conclusions} concludes the paper.
\vskip2mm
\textit{Notations:} Boldface letters and upper-case letters denote column vectors and matrices, respectively.
The $i$th element of vector $\xv$ is denoted $x_i$.
The notation 
$\xv^{H}$ denotes the Hermitian transpose of $\xv$.

\section{System model and fundamental limitations}
\label{sec:model_and_limits}
\begin{figure}[H]
	\centering
	\includegraphics{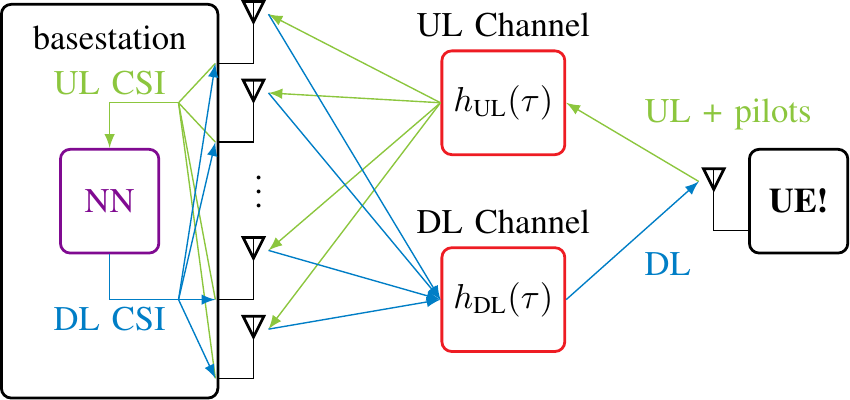}
	\caption{The \ac{DL} \ac{CSI} extrapolation is based on a \ac{NN} trained using \ac{UL} \ac{CSI} observations.} 
	\label{fig:System-Model-Chan-Est}
\end{figure}

Fig.~\ref{fig:System-Model-Chan-Est} depicts the fundamental problem we tackle in this work. 
The task of the \ac{BS} is to estimate the \ac{DL} \ac{CSI} $h_{\text{DL}}(\tau)$ only by observing the \ac{UL} \ac{CSI} $h_{\text{UL}}(\tau)$ (\ac{UL} pilots) as shown in Fig.~\ref{fig:System-Model-Chan-Est}.\footnote{We consider a scenario where the \ac{BS} has multiple antennas, while each \ac{UE} has a single antenna. Extensions to \acp{UE} with multiple antennas are straightforward.}
Obviously, the success of such an approach relies on whether there is an underlying physical relation between the considered frequency bands; thus, in Section~\ref{sec:siso} and Section \ref{sec:mimo}, respectively, we demonstrate the validity of this assumption for both simulated and measured scenarios.

\subsection{Approximating the channel's impulse response}

As shown in Fig.~\ref{fig:System-Model-Chan-Est}, we consider a wireless channel between two radio transceivers, referred to as \acf{UE} and \acf{BS}, which are both equipped with a single antenna.
The channel impulse response $h(\tau)$ can be written as
\begin{align}\label{eq:h}
h(\tau) = \sum_{p=1}^P a_p \delta(\tau-\tau_p)
\end{align}
where $a_p\in \mathbb R_+$ and $\tau_p\in \mathbb R_+$ for $p=1,\dots,P$ are the attenuation and propagation delay of the $p$th path, respectively.
The channel impulse response is reciprocal, i.e., it characterizes the channel between both transceivers in each directions.
Both transceivers communicate using an \ac{FDD} scheme, where one frequency band is used for \ac{UL} transmissions from the \ac{UE} to the \ac{BS}, while another frequency band is used for \ac{DL} transmissions from the \ac{BS} to the \ac{UE}.
Both frequency bands are of bandwidth $W$ and have center frequencies $f_\text{UL}$ and $f_\text{DL}$, respectively.
In the following we assume that \eqref{eq:h} is valid over a large frequency range $\left[ \min(f_\text{UL},f_\text{DL})-W/2, \max(f_\text{UL},f_\text{DL})+W/2 \right]$, sufficient to contain both the \ac{UL} and \ac{DL} frequency bands.

Following \cite{tse2005fundamentals}, the equivalent discrete time baseband model of the channel $h(\tau)$ over a frequency band of bandwidth $W$ and center frequency $f_c$ is given as 
\begin{equation}\label{eq:corr_fundamental}
h_{\ell} = \sum_{p=1}^{P} a_{p} e^{-j 2\pi f_c \tau_p} \text{sinc}\left( \ell -\tau_pW\right) \quad \ell=0,\dots,L-1
\end{equation}
where $\text{sinc}(x)=\frac{\sin(\pi x)}{\pi x}$ is the normalized sinc-function and $L$ is the largest integer for which $|h_L|\ge \varepsilon$, for some threshold $\varepsilon > 0$.
Evaluating \eqref{eq:corr_fundamental} for $f_c = f_\text{UL}$ and $f_c = f_\text{DL}$, we obtain the equivalent discrete time baseband \ac{UL} and \ac{DL} channels $h_{\ell,\text{UL}}\in\mathbb C$ and $h_{\ell,\text{DL}} \in\mathbb C$, respectively.

Due to the finite sampling rate $W$, it is generally impossible to estimate the exact values of $\tau_p$ from $h_{\ell,\text{UL}}$ or $h_{\ell,\text{DL}}$ for any arbitrary $f_\text{UL}$ to $f_\text{DL}$, e.g., from \SI{1}{\giga \hertz} to \SI{6}{\giga \hertz}.
It is, therefore, also impossible to compute the exact values of $h_{\ell,\text{DL}}$ from $h_{\ell,\text{UL}}$ and vice versa.
However, a \ac{NN} can predict an \emph{approximation} of $h_{\ell,\text{DL}}$ from $h_{\ell,\text{UL}}$ (or vice versa) for a given radio environment at an adjacent frequency band.

We denote by $\hv_{\text{UL}}\in\mathbb C^{N_{\text{sub}}}$ and $\hv_{\text{DL}}\in\mathbb C^{N_{\text{sub}}}$ the \ac{CSI} vectors over subcarriers of an \ac{OFDM} system.

\subsection{Metrics}\label{sec:Metrics}

Throughout this work, three different performance metrics are used, each of which coming with its own strengths and weaknesses, as will be discussed next.

\subsubsection{Normalized mean squared error (NMSE)} \acused{NMSE}

We use the \ac{NMSE} as the loss function during training given as

\begin{align}
\text{NMSE} = \mathbb{E}\left[ \frac{\Vert \hv_\text{DL} - \hat{\hv}_\text{DL}\Vert^2_2}{\Vert \hv_\text{DL}\Vert^2_2} \right].
\label{eq:nmse}
\end{align}
While the \ac{MSE} is a standard metric for regression tasks, our task requires a normalization such that the performance is independent of the path loss, i.e., $\mathbb{E} \left[ \left| \hv_\text{DL} \right|^2 \right] = 1$, which would otherwise suggest misleading results.
The normalization forces the \ac{NN} to also focus on channel coefficient characteristics for distant \ac{UE} positions, where $\left| \hv_{\text{DL}} \right|^2$ is comparatively small.
By only training on the \ac{MSE} without normalization, the \ac{NN} tends to learn to output a random near-zero prediction for input \ac{CSI} where $\mathbb{E} \left[ \left| \hv_\text{UL} \right|^2 \right]$ is small.
However, the drawback of the \ac{NMSE} is that its interpretation is less intuitive in how the \ac{NMSE} relates to the achieved performance and, thus, we, rather, consider the correlation coefficient for most evaluations.

\subsubsection{Correlation coefficient}

We use the \textit{correlation coefficient} $\delta_{h}$ as an alternative performance metric, allowing a more intuitive interpretation.
The correlation coefficient is defined as 
\begin{align}
\delta_{h}= \mathbb{E}\left[\frac{\left\Vert \hv_{\text{DL}} \hat{\hv}_{\text{DL}}^{H}\right\Vert_\text{2}}{\left\Vert \hv_{\text{DL}} \right\Vert_{\text{2}}\Vert {\hat{\hv}_{\text{DL}}}\Vert _{\text{2}}}\right] =\mathbb{E}\left[ \frac{\sum_{k=0}^{N_\text{sub}}{\Vert h_{k,\text{DL}} \hat{h}_{k,\text{DL}} \Vert_\text{2}}}{\left\Vert \hv_{\text{DL}} \right\Vert_{\text{2}}\Vert {\hat{\hv}_{\text{DL}}}\Vert_{2}} \right].
\label{eq:ChannelCorrDef}
\end{align}

Its main advantage is that it provides a very intuitive result, as the range is bounded by the perfect match between predicted and actual channel (i.e., $\delta_h=1$) and an uncorrelated scenario (i.e., $\delta_h=0$).
Note that, a common phase rotation for all elements of the prediction does not affect this metric. However, a constant phase offset is inherently compensated by the underlying \ac{OFDM} scheme.
In our experiments we observed, in contrast to \ac{NMSE}, that the correlation coefficient is not a suitable loss function. It converged quickly to a local minimum with unsatisfactory performance.

\subsubsection{Bit error rate (BER)}\acused{BER}
Despite the fact that the \ac{NMSE} and correlation coefficient provide good metrics for comparing the similarity of the predicted \ac{CSI}, it does not tell us anything about the resulting \ac{BER} performance.
Thus, we also evaluate the (uncoded) \ac{BER} when operating the system with the predicted channel coefficients.
This means we precode/equalize \ac{QPSK} symbols based on predicted \ac{CSI} by the \ac{NN}, and then evaluate on a channel that uses the actual \ac{CSI}.
While the \ac{BER} metric also provides a good insight into the \ac{NN}'s performance, its drawback is that, unfortunately, it cannot be directly embedded as loss function, as it requires intensive Monte-Carlo simulations per prediction, and, moreover, is not differentiable.
One could possibly use soft information, i.e., cross-entropy if needed.

\section{A primer on Deep Learning and Datasets}
\label{sec:deep_learning}
We provide a brief introduction to deep learning with the aim of clarifying notation and terminology used throughout this work.
However, we refer the interested reader to \cite{goodfellow2016} for further details on deep learning in general.

An \acf{NN} consists of weights $\boldsymbol{\theta}$, within a certain layer structure, that define a mapping 
\begin{align}
\hat{f}(\xv;\boldsymbol{\theta}): \RR^n \mapsto \RR^k
\end{align}
of an input vector $\xv\in\RR^n$ to an output vector $\hat{\yv}\in\RR^k$.
The weights $\boldsymbol{\theta}$ determine the \ac{NN}'s behavior, and the process of finding good values for $\boldsymbol{\theta}$ from data to achieve a desired behavior is generally described as ``deep learning''.
In our case the input vector $\xv$ is the \ac{CSI} of one channel (subcarriers) at a frequency centered around $f_\text{UL}$ and the desired behavior of the \ac{NN} is to output a prediction $\hat{\yv}$ of the \ac{CSI} of another channel at another frequency band centered around $f_\text{DL}$.
This is called a regression task and we can use well-established algorithms to fit the weights to our datasets such that they minimize a certain loss metric.
A single complex-valued number is split into two consecutive real-valued numbers and used as input for the \ac{NN} and, vice versa, at the output.

We start training with small mini-batches containing only $16$ samples and increase the batchsize during the process stepwise up to $512$ to obtain more fine-grained weight updates.
During training, we also add \ac{AWGN} as regularization to the training samples to prevent overfitting.

Intuitively, the optimal training \ac{SNR} is a trade-off between high noise power, i.e., \emph{learning robustness to noisy data} and noiseless samples, i.e., \emph{learning the underlying (deterministic) channel transfer function} \cite{gruber2017}.

\subsection{Dataset}
We can easily generate large quantities of simulated or measured channel realizations on both frequency bands at $f_\text{UL}$ and $f_\text{DL}$, respectively. Our datasets contain labeled samples, i.e., a sample is a tuple of input $\xv = \hv_\text{UL}$ and label $\yv = \hv_\text{DL}$.
Our dataset contains $N$ samples denoted by $\{(\hv_{\text{UL},1},\hv_{\text{DL},1}),\dots,(\hv_{\text{UL},N},\hv_{\text{DL},N})\}$.
In all our simulations and measurements we use an \ac{OFDM} scheme with $N_{sub}=1024$ subcarriers.
Due to guard-bands, we effectively use 922 out of those $1024$ subcarriers, resulting in a \ac{CSI} vector of 922 complex channel coefficients.
We reshape those coefficients to their real-valued real and imaginary parts and also use multiple antennas in the \ac{MIMO} scenario.
Thus, the dimensionality of both the input and the label tensor is $M \times N_{\text{sub}} \times 2$ (nb.\ antennas, nb.\ subcarriers, real/imaginary part).
All datasets are randomly split into a train set, containing 90\%, and a test set, containing 10\% of all samples.

\subsection{Convolutional Layers}
Our final \ac{NN} structure mainly consists of \ac{CNN} layers, reducing the total amount of trainable parameters by order of magnitudes.
Since neighboring \ac{OFDM} channel coefficients in frequency domain are correlated within the coherence bandwidth, we figured that convolutional layers are perfectly suited to extract latent information.
Thus, we use multiple two-dimensional convolutional layers with different kernel sizes in our final \ac{NN} structure, which is similar to established models in computer vision like VGG \cite{simonyan2014very}.
Furthermore, additional pooling layers after each \ac{CNN} layer are used to reduce the dimensionality of the input tensor.
As final output we use a dense layer with linear activation to create the channel coefficients of the \ac{DL} in the desired output shape.

\section{\ac{SISO} \ac{UL}-\ac{DL} channel prediction}
\label{sec:siso}

We first focus on the \ac{SISO} scenario to illustrate the differences between the previously described performance metrics, and show the viability of this concept.
However, it is important to realize that a solution for the \ac{SISO} scenario directly provides a \emph{naive} \ac{MIMO} solution (by applying $M$ independent estimators).

\subsection{Basic Example: \acf{LoS} Model}
\label{sec:siso-LoS}
If we consider a pure \ac{LoS} scenario with distance dependent pathloss, the \ac{UL} and \ac{DL} channel impulse responses for a distance $d$ between both transceivers are given by 
\begin{align}
\label{eq:UplinkDownlinkEq}
h_\text{UL}(d) &= \left(\frac{c}{4\pi f_{\text{UL}}d}\right)^\beta e^{-j2\pi f_\text{UL} \frac{d}{c}}\\
h_\text{DL}(d) &= \left(\frac{c}{4\pi f_{\text{DL}}d}\right)^\beta e^{-j2\pi f_\text{DL} \frac{d}{c}}
\end{align}
where $\beta>2$ is a pathloss exponent and $c$ is the speed of light.
In this scenario, we do not use \ac{OFDM} channel coefficients, since there is only a single \ac{LoS} path, resulting in a system which is already frequency flat.
Therefore, we only consider one channel coefficient for each $f_\text{UL}$ and $f_\text{DL}$, respectively.
We also denote the \ac{UL}/\ac{DL} band separation as $\Delta f = f_\text{DL} - f_\text{UL}$ in the following.
Since the underlying channel function is known in this example, there are several approaches to tackle the task of estimating $h_\text{DL}$ from $h_\text{UL}$.

\subsubsection{Classic analytical channel estimation}

It is apparent from \eqref{eq:UplinkDownlinkEq} that the absolute value of $|h_\text{UL}|$ is only dependent on $d$.
Thus, the only difficulty is to predict the distance $d$ of $h_\text{UL}$ based on the magnitude of $h_\text{UL}$. 
Since $d=\frac{c}{4\pi  f_\text{UL} |h_\text{UL}(d)|^{\frac1\beta}}$, we have 
\begin{align}
h_\text{DL}(d) = \left(\frac{f_{\text{UL}}}{f_{\text{DL}}}\right)^\beta |h_\text{UL}(d)| \exp \left( -j \frac{f_{\text{DL}}}{2f_{\text{UL}}|h_\text{UL}(d)|^{1/\beta}} \right). \label{eq:los-uldl}
\end{align}
This means that a \ac{NN} only needs to learn a mapping of the form $f(x) = x e^{-jK|x|^{-\frac1\beta}}$
(and the distance $d$) from the observations, which is clearly possible provided that enough training samples $\{(\hv_{\text{UL},1},\hv_{\text{DL},1}),\dots,(\hv_{\text{UL},N},\hv_{\text{DL},N})\}$ are available.
Note that this approach only works as a baseline for the pure \ac{LoS} scenario.

\subsubsection{Wiener filter estimation}
A classic approach to estimate \ac{CSI} in time and frequency domain is the Wiener filter (also referred to as \ac{LMMSE} estimator) \cite{780524}.
The Wiener filter coefficients are defined as
\begin{equation}
\cv_\text{LMMSE} = \left(\Rm_\text{UL,UL} + \sigma^2 \bf{I} \right)^{-1} \mathbb{E}\left[\hv_\text{UL} \hv_\text{DL}^\text{H} \right]
\end{equation} 
with noise variance $\sigma^2$ (complex noise samples) and
\begin{equation}
\Rm_\text{UL,UL} = \mathbb{E}\left[\hv_\text{UL} \hv_\text{UL}^\text{H} \right].
\end{equation}
To estimate unknown \ac{CSI}, we match the closest known validation $\hv_{\text{UL}}$ point to the current test point $\hv'_{\text{UL}}$ by using the correlation and by taking its filter coefficients $\cv'_\text{LMMSE}$ to estimate
\begin{equation}
\hat{\hv}_\text{DL} = {\cv^{'\text{H}}_\text{LMMSE}} \hv'_\text{UL}.
\end{equation}

\subsubsection{NN-prediction approach}
The fact that there exists a function \eqref{eq:los-uldl} that analytically solves this toy case, the universal approximation theorem \cite{cybenko1989approximation} tells us that there exists a \ac{NN} that can approximate, or \emph{learn}, this function arbitrary well.
We therefore use a basic \ac{NN} structure (shown in Table~\ref{tab:NN-LoS-SISO}) based on feed-forward dense layers to solve this task.
\begin{table}[H]
\centering
\caption{\ac{NN} architecture for \ac{LoS} experiments}
\begin{tabular}{c|c|c}
Layers:                      & Trainable parameters      & Output dimensions           \\
\hline                                                                       
Input                  & 0               & 2 (Re/Im)   \\
Dense                  & \num{384}       & 128                        \\
Dense                  & \num{33024}     & 256                        \\
Dense                  & \num{263168}    & 1024                           \\
Dense                  & \num{26400}     & 256                           \\
Dense                  & \num{32896}     & 128                        \\
Dense                  & \num{258}       & 2 (Re/Im)                 \\
\end{tabular}
\label{tab:NN-LoS-SISO}
\end{table}

The \ac{NN} receives a single channel coefficient $h_\text{UL}$ as input and predicts the corresponding single channel coefficient $h_\text{DL}$ as output.
We train this \ac{NN} with samples created according to \eqref{eq:UplinkDownlinkEq}, where the \ac{UE} is randomly positioned within a radius ranging from \SI{100}{\metre} to \SI{200}{\metre} around the \ac{BS}.
Also, the results are inferred on random positions within this range.

\vskip2mm
We evaluate all three aforementioned approaches with 5,000 randomly positioned samples (\acp{UE}) based on the \ac{LoS} pathloss model \eqref{eq:UplinkDownlinkEq} and test all approaches on predicting the correct $h_\text{DL}$ from the given $h_\text{UL}$.
Then, we take the previously described \ac{BER} metric with predictions $\hat{h}_\text{DL}$ to precode/equalize \ac{QPSK} symbols transmitted over the actual channel characterized by $h_\text{DL}$.

\begin{figure}[H]
	\centering
	\includegraphics{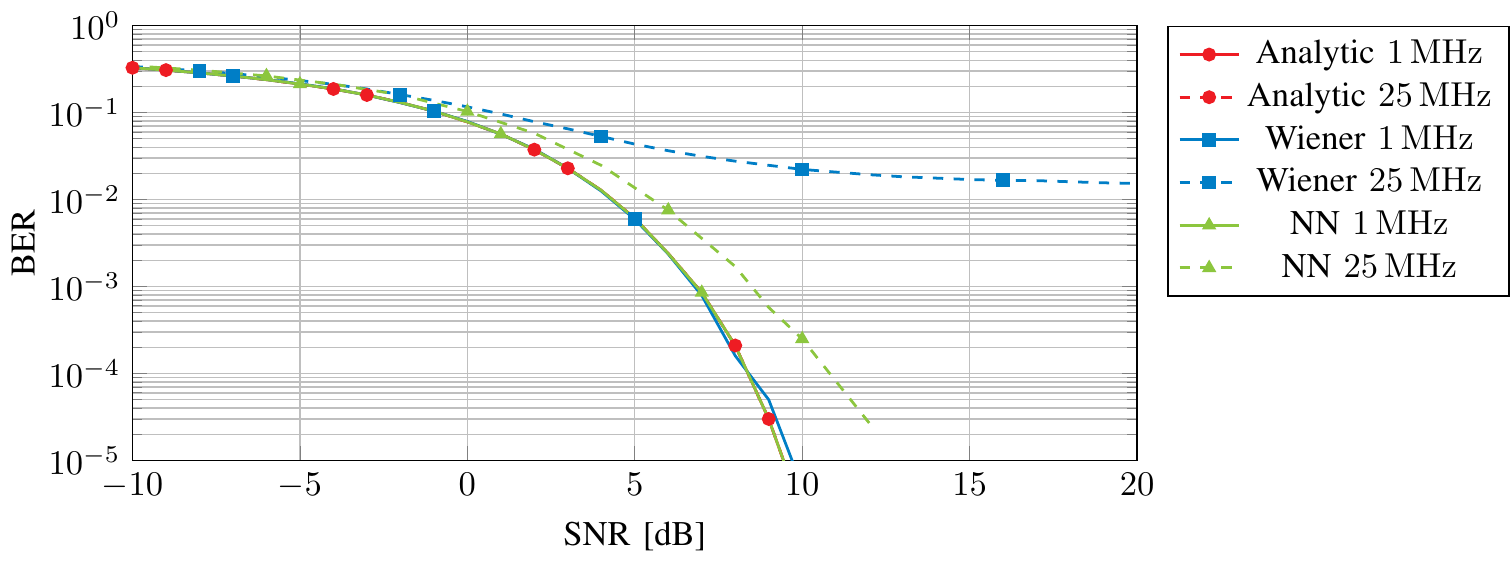}
	\caption{\ac{BER} curves of different estimation approaches in the \ac{LoS} scenario; the \ac{NN} is trained with 4,500 samples.} 
	\label{fig:Comp-Est-BER-SISO_LoS}
\end{figure}

Fig~\ref{fig:Comp-Est-BER-SISO_LoS} shows the results for a \ac{UL}/\ac{DL} band separation of $\Delta f= f_\text{DL}- f_\text{UL} =\SI{1}{\mega \hertz}$ and $\Delta f= \SI{25}{\mega \hertz}$, respectively.
As expected, the analytical solution leads to optimal results, hence coincides with the uncoded \ac{QPSK} \ac{BER} curve (the same holds for the Wiener filter at $\Delta f= \SI{1}{\mega \hertz}$).
However, the more interesting observation is that the Wiener filter only works well for $\Delta f= \SI{1}{\mega \hertz}$, but not for $\Delta f=\SI{25}{\mega \hertz}$.
Intuitively, this can be explained by the fact that the Wiener filter needs to extrapolate the \ac{DL} \ac{CSI} from only one specific \ac{UL} \ac{CSI} sample depending on the actual position.
The autocorrelation matrix does not contain the spatial dependencies and, thus, the classical Wiener filter approach \emph{fails} for most tasks.
Unfortunately, to the best of our knowledge, this is the best estimator for extrapolation in frequency domain (without any further channel sparsity assumption).
The \ac{NN}, on the other hand, achieves optimal results for $\Delta f=\SI{1}{\mega \hertz}$, however, shows some degradation for $\Delta f=\SI{25}{\mega \hertz}$.
This degradation can easily be reduced by supplying more training samples.
By limiting the amount of training samples to only 4,500 for both the $\Delta f= \SI{1}{\mega \hertz}$ separation and the $\Delta f= \SI{25}{\mega \hertz}$ separation, we can see that predictions on a close-by frequency band ($\Delta f= \SI{1}{\mega \hertz}$) are \emph{easier} for the \ac{NN}.
We also achieved optimal prediction results for $\Delta f= \SI{25}{\mega \hertz}$ in this simple \ac{LoS} toy case scenario after training the \ac{NN} with $\geq$10,000 samples.
Contrary to the analytical solution, the \ac{NN} approach is completely independent of knowing the underlying channel model, whereas the analytical solution only works for this simple \ac{LoS} model.

\begin{figure}[H]
	\centering
	\includegraphics{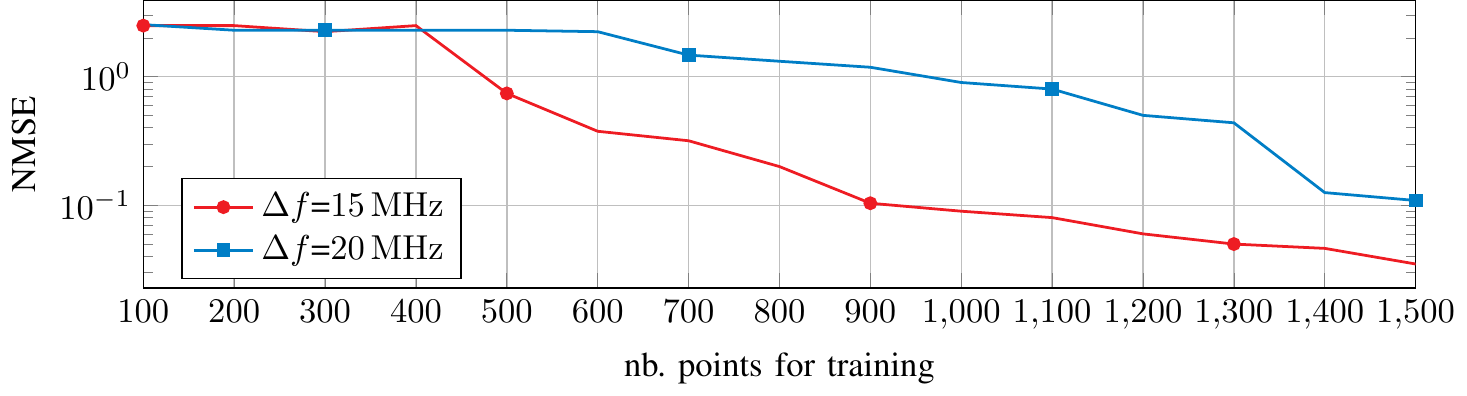}
	\caption{NMSE versus number of training points and different frequency band separations $\Delta f$.}
	\label{fig:Influe-NbPoints-Deltaf}
\end{figure}
Fig.~\ref{fig:Influe-NbPoints-Deltaf} shows the influence of different amounts of training samples for $\Delta f=\SI{15}{\mega \hertz}$ and $\Delta f=\SI{25}{\mega \hertz}$, respectively.
It takes about 400 to 600 different samples during training for the \ac{NN} to start to generalize.
Obviously, a larger frequency band separation $\Delta f$ between \ac{UL} and \ac{DL} also requires a higher number of training samples to sufficiently solve the task, as the relation between input and label appears to be more \emph{random} from the \acp{NN}'s point-of-view.
Intuitively, this can be explained due to periodicity of the phase rotation in \eqref{eq:UplinkDownlinkEq}, causing the same input phase to be projected onto more than one output phase.
Therefore, the single most relevant variable, that is, the amplitude, needs to be estimated more precisely for larger band separation $\Delta f$.

\subsection{Simulated Results: Standardized Channel Models}\label{sec:siso_sim_soph}

Next, we evaluate \ac{NN}-based channel prediction in a more realistic setting using standardized channel models provided by the Quadriga framework \cite{Jaeckel2014}, such as the \emph{Winner II}, \emph{3GPP} and \emph{Berlin} model.

\begin{figure}[H]
	\centering
	\includegraphics{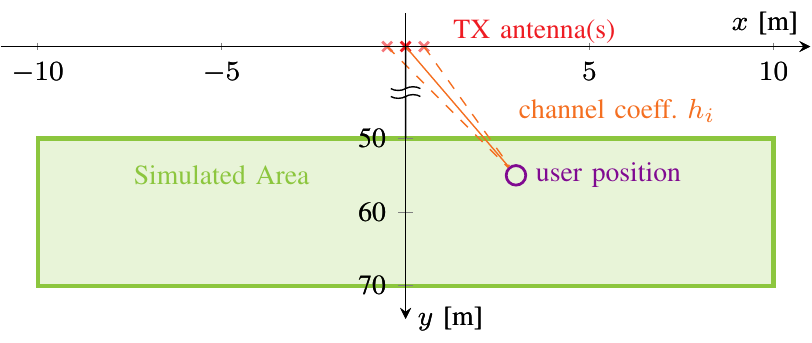}
	\caption{Spatial simulation area for different channel models.} 
	\label{fig:SISO-Quad-SimArea}
\end{figure}

The simulation setup is depicted in Fig.~\ref{fig:SISO-Quad-SimArea}.
We use the channel models to simulate the \ac{CSI} at $f_\text{UL}$ and $f_\text{DL}$ between a fixed \ac{BS} and a \ac{UE} at random spatial positions within the same environment.
An area of $400m^2$ is simulated with an average distance per spatial \ac{UE} position of $\sim 12.5cm$ resulting in 22,500 samples.
Per \ac{UE} position, we estimate the channel for two different frequency band separations ($\Delta f=\SI{25}{\mega \hertz}$, $\Delta f=\SI{50}{\mega \hertz}$) between $f_\text{UL}$ and $f_\text{DL}$.
For this, we use an \ac{OFDM} system with 1024 subcarriers and a \ac{CP} length of 256 symbols with two dipole antennas at frequency bands centered around \SI{1.25}{\giga \hertz} and \SI{1.275}{\giga \hertz} (\SI{1.3}{\giga \hertz} for $\Delta f=\SI{50}{\mega \hertz}$, respectively) with a bandwidth of $B=\SI{20}{\mega \hertz}$ for both the UL and DL frequency band.

Note that it is necessary to enable Quadriga's \emph{spatial consistency} parameter to ensure that the simulation environment, i.e., positions of scatterers, does not change when the \ac{UE} changes its position.
Random realizations of different environments would render the channel prediction task impossible, as the \ac{NN} would not be able to learn an underlying physical channel scattering behavior.
For all \ac{SISO} results, the antenna geometry does not need to be considered and, thus, we only use 1-D convolutional layers in our \acp{NN}.
The hyperparameters of this \ac{SISO} \ac{NN} structure, as described in Tab.~\ref{tab:SISO-Quadriga-NNLayout}, was trained at an \ac{SNR}=$\Vert \hv \Vert^2_\text{2}/2\sigma^2$  of \SI{10}{\decibel}, where $\sigma^2$ is the noise variance.

\begin{table}[H]
\centering
\caption{\ac{NN} architecture for \ac{SISO} experiments}
\begin{tabular}{c|c|c}
Layers:                      & Trainable parameters      & Output dimensions           \\
\hline                                                                       
Input                  & 0              & 1024 x 2 (Re/Im)   \\
Conv1D                 & \num{416}      & 1024 x 32                  \\
Average Pooling         & \num{0}        & 256 x 32                  \\
Conv1D                 & \num{3088}     & 256 x16                \\
Average Pooling         & \num{0}        & 64  x16                   \\
Flatten                & \num{0}        & 1024                           \\
Dense                  & \num{131200}   & 128                          \\
Dense                  & \num{16512}    & 128                        \\
Dense                  & \num{16512}    & 128                        \\
Dense                  & \num{264192}   & 2048                        \\
Reshape                & \num{0}        & 1024 x2  (Re/Im)   \\
\end{tabular}
\label{tab:SISO-Quadriga-NNLayout}
\end{table}

\begin{figure}[H]
	\centering
	\includegraphics{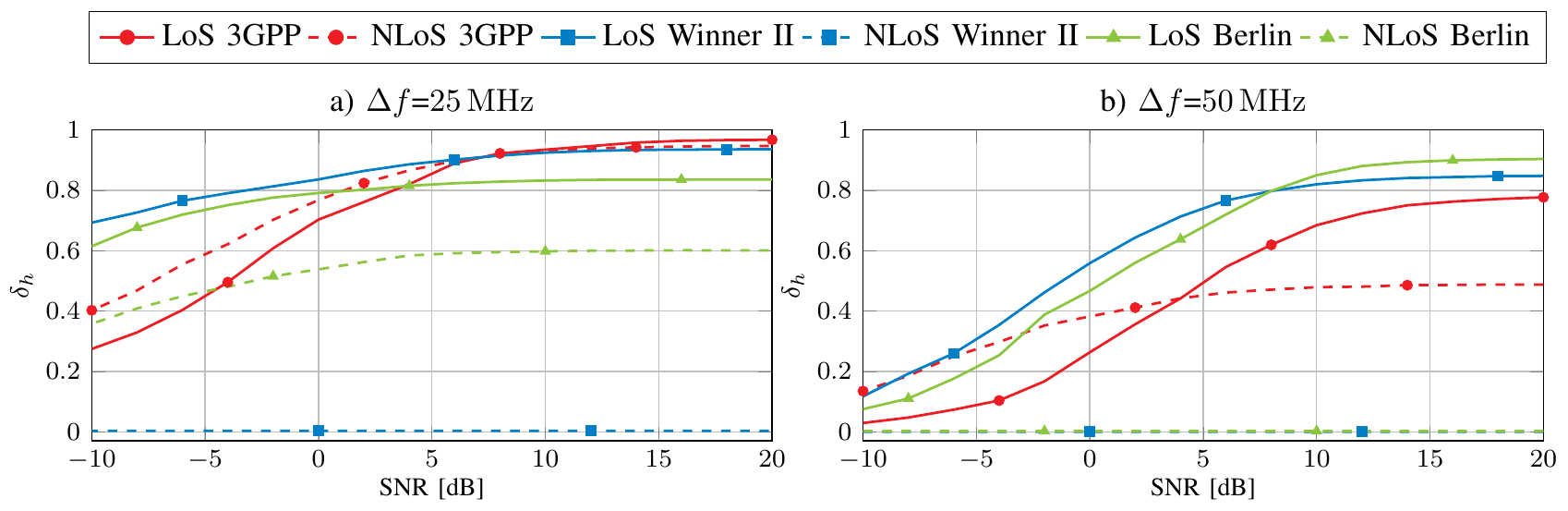}
	\caption{\ac{NN}-based \ac{CSI}-prediction: correlation coefficient versus \ac{SNR} with different \ac{SISO} channel models at band separations $\Delta f=\SI{25}{\mega \hertz}$ (a) and $\Delta f=\SI{50}{\mega \hertz}$ (b).} 
	\label{fig:SISO-Corre-25MHz}
\end{figure}
Fig.~\ref{fig:SISO-Corre-25MHz} shows the correlation coefficient $\delta_h$ for a frequency band separation of $\Delta f=\SI{25}{\mega \hertz}$ (left) and  $\Delta f=\SI{50}{\mega \hertz}$ (right) between the \ac{UL} and \ac{DL} versus the \ac{SNR}.
As can be seen, the \ac{NN} estimator (see Table~\ref{tab:SISO-Quadriga-NNLayout}) is able to predict the \ac{CSI} for these standardized channel models, and the \ac{NN} can generalize over a wide range of different \acp{SNR}.
In the \ac{LoS} scenario, all three different channel models lead to a reasonable performance ($\delta_h$ $\geq$ 0.8) by the \ac{NN}.
Simulations on spectral efficiency (shown later in Section~\ref{sec:mimo}) indicate that a correlation coefficient of $\delta_h=0.8$ results in an \ac{MRT} sum-rate performance loss of less than 10\%. 
For the \emph{3GPP} model, the \acp{NN} performance is very similar for both the \ac{LoS} and \ac{NLoS} scenario in terms of the correlation metric.
Although the performance is worse for the \ac{NLoS} scenario with the \emph{Berlin} model, a  generalization of the \ac{NN} can still be observed, resulting in a correlation coefficient of $\delta_h\sim0.6$.
The prediction does not work at all for the \ac{NLoS} scenario with the \emph{Winner II} model.
As already mentioned concerning the \emph{spatial consistency} parameter, the variations among the results of the three models can be explained by the differences between the channel models themselves:
\begin{itemize}
\item The 3GPP model uses consistency in both the spatial domain and the frequency domain in accordance to  3GPP 38.901 v14.1.0.
Therefore, the \ac{NN} can predict between close frequencies. 
\item Since the Winner II model is not based on this feature \cite{Jaeckel2014}, the \ac{NN} can only learn the environment in a more or less frequency flat scenario (\ac{LoS}).
\item Although the \ac{NLoS} Berlin model is based on measurements (from which parameters were extracted, which were then used to re-simulate the measurement results) the model relies mainly on statistics rather than on actual environments (scatterers).
Thus, an underlying frequency dependency does probably not exist in this model.
\end{itemize}

As expected, when the frequency band separation increases to $\Delta f$=\SI{50}{\mega \hertz}, the correlation coefficient decreases.
Although the \ac{NN}-based prediction still works in the \ac{LoS} scenarios, no meaningful prediction can be provided for the \ac{NLoS} scenario due to a weaker frequency dependency between \ac{UL} and \ac{DL}.
However, a larger training dataset may further improve the performance.
Note again that the Wiener filter-based approach fails for all investigated scenarios.

\begin{figure}[H]
	\centering
	\includegraphics{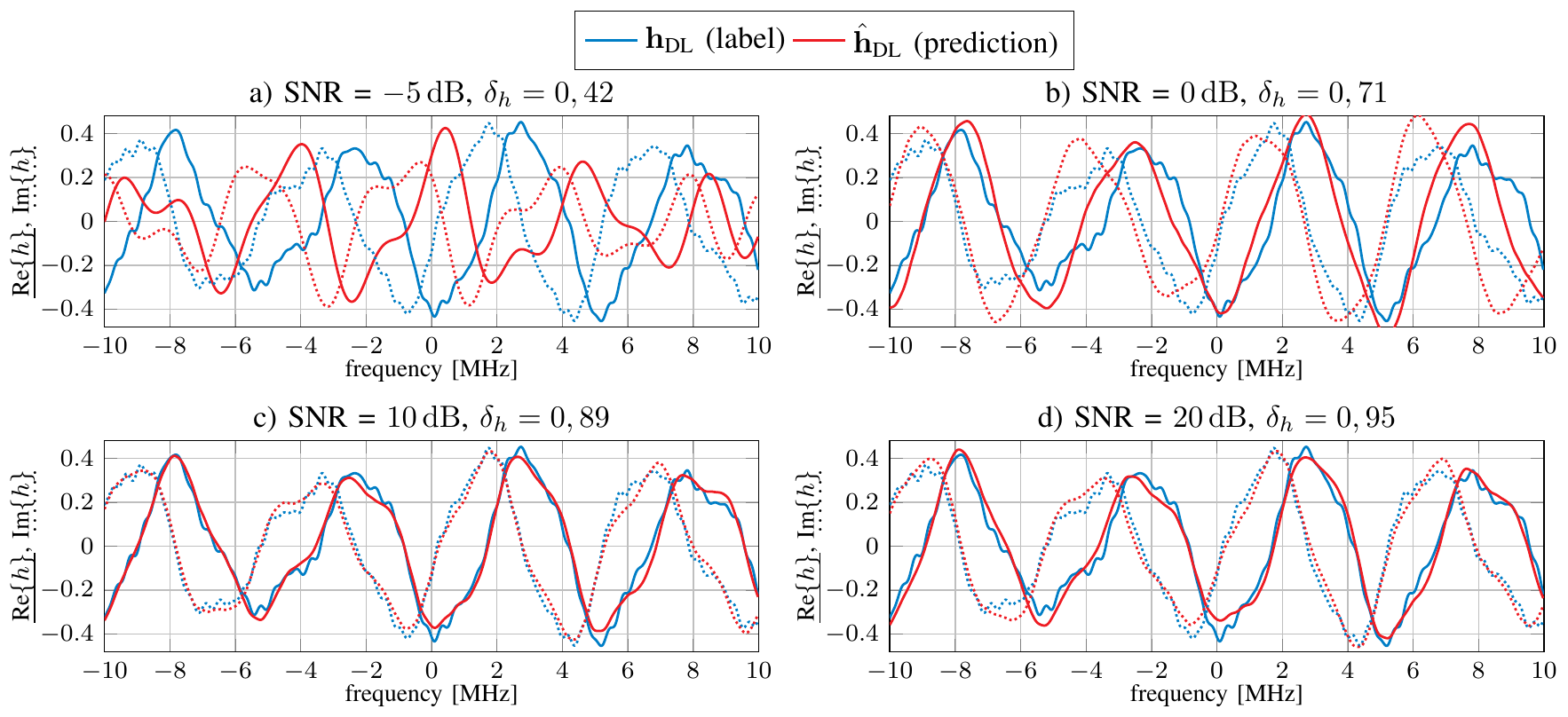}
	\caption{Illustration of \ac{NN}-based channel prediction for varying \acp{SNR}, 3GPP 38.901 \ac{LoS} channel model.}
	\label{fig:Illust-Prediction-of-simChans}
\end{figure}
To better illustrate the impact of the correlation coefficent, Fig.~\ref{fig:Illust-Prediction-of-simChans} shows a visualization of the prediction outcome and the corresponding ground truth for the \emph{3GPP} 38.901 model versus different \ac{SNR}, while the \ac{NN} was trained at SNR = \SI{10}{\decibel}. 
Note that the channel can be accurately predicted between \ac{UL} and \ac{DL} frequency bands for positive values of the SNR (in dB).
Thus, as a first major result of this paper, we show that the proposed \ac{NN} architecture can predict the \ac{DL} \ac{CSI} for the \ac{SISO} scenario.\footnote{We want to emphasize that such a \ac{SISO} result directly enables naive \ac{MIMO} implementation by using $M$ independent estimators.}

\subsection{Measurement data: SISO}\label{sec:Meas-SISO}

Although the previously introduced channel models provide quite realistic test cases, and their properties have been verified through measurements by multiple independent research groups, the question whether \ac{DL} channel prediction works on \emph{actually measured} channels still remains open.
One should keep in mind that most channel models are built for system performance evaluation where exact position-dependent precision is typically unnecessary. 
Further, we expect many hardware impairments in a practical system, which are difficult to cover in model-based approaches.
To address these issues, we evaluate the \ac{NN} prediction performance on measured channels, as described next. 

\begin{figure}[]
	\centering
	\includegraphics[width=0.5\textwidth]{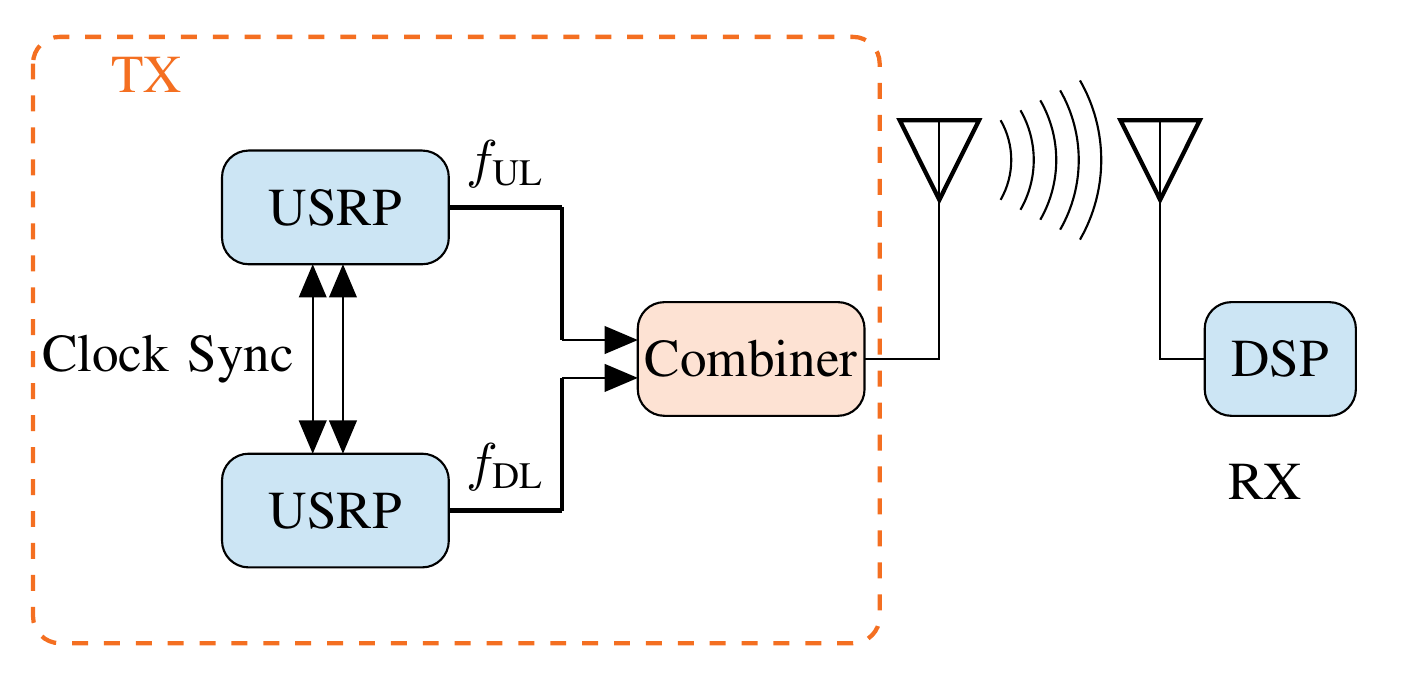}
	\caption{Setup for the \ac{SISO} link measurement.}
	\label{fig:MeasurementSetupSISOLink}
\end{figure} 

To measure any combination of \ac{UL} and \ac{DL} frequencies, two \acp{USRP} (B200 + B210) \cite{USRP} are synchronized and combined on a single antenna as shown in Fig.~\ref{fig:MeasurementSetupSISOLink}.
A single antenna is needed to ensure the exact same distance $d$ between the transceivers.
Typically, \ac{FDD} systems also use the same antenna (in this case a horizontally polarized dipole) for \ac{UL} and \ac{DL} transmissions.
The \acp{USRP}s transmit at \SI{1.25}{\giga \hertz}, \SI{1.275}{\giga \hertz} and \SI{1.3}{\giga \hertz}, respectively, with \SI{20}{\mega \hertz} bandwidth each.
The radio signal is transmitted over the channel, received with the horizontally polarized dipole antenna and then sampled at the carrier frequency by a wideband digital oscilloscope.
\ac{UL} and \ac{DL} transmissions are then separated by an offline signal processing step (filtering).
An extension to multiple (\ac{MIMO}) receiver antennas is straightforward (see \cite[daughterboard concept]{Arnold2018}).

\begin{figure}[H]
     \centering
     \subfloat[Indoor measurement scenario]{\includegraphics{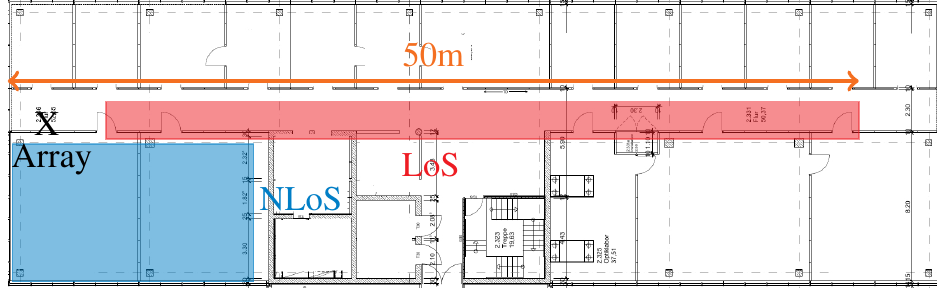}}
     \subfloat[Outdoor measurement scenario]{\includegraphics{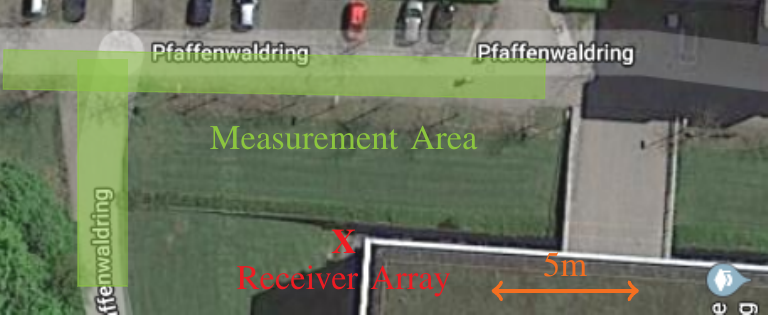}}
     \caption{Measurement scenarios: a) indoor b) outdoor.}
     \label{fig:MeasurementScenarios}
\end{figure}

To evaluate the performance in different scenarios, we distinguish between indoor and outdoor environments (see Fig.~\ref{fig:MeasurementScenarios}).
Note that, the same scenarios are also used for the \ac{MIMO} setup later in Section~\ref{sec:mimo}.
The indoor measurements include an \ac{LoS} environment, consisting of our institute's hallway, and an \ac{NLoS} environment, represented by typical office space.
Both indoor measurements were conducted within a static environment (at night), i.e., the environment did not change significantly during the measurements.
The covered area is about \SI{200}{\metre \squared} in the \ac{LoS} scenario and about \SI{80}{\metre \squared} in the \ac{NLoS} scenario.
For the outdoor measurement, a typical Rician fading scenario can be expected, in a campus environment with tall buildings surrounding the measured area of about \SI{250}{\metre \squared}.
We acquired 4,000 measurement samples per scenario.
All positions were measured in a meander-like path, however, the samples were randomly permuted during training, i.e., training and validation data were randomly selected from the entire area (90\% training and 10\% validation).
\begin{figure}[H]
	\centering
	\includegraphics{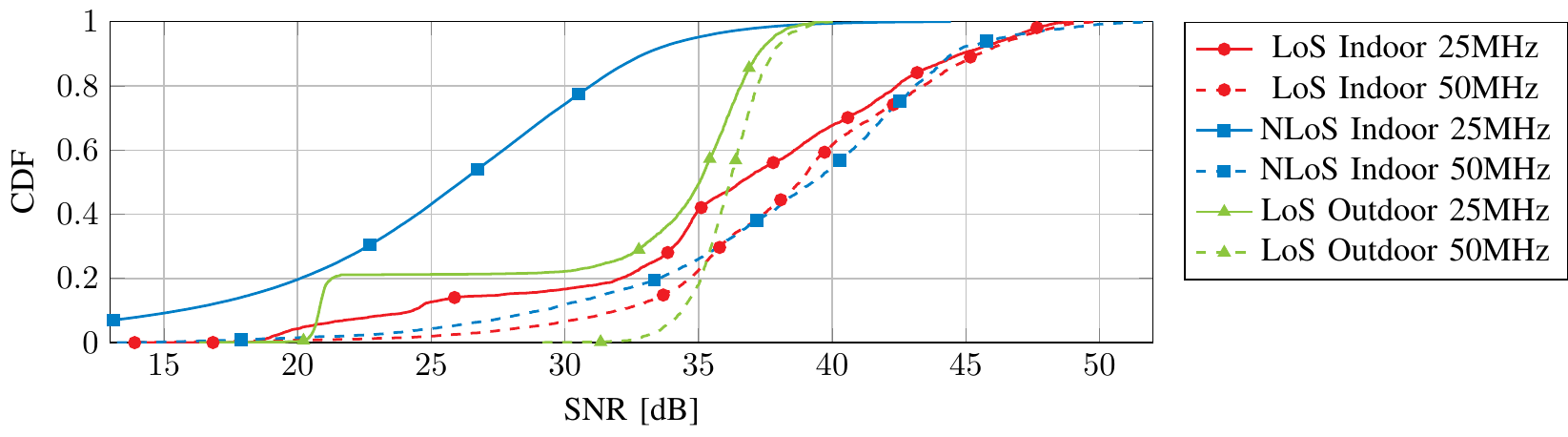}
	\caption{\ac{CDF} of the received \ac{SNR} for different measurement scenarios.}
	\label{fig:SISO-Meas-SNR}
\end{figure}

To verify the measurement data, Fig.~\ref{fig:SISO-Meas-SNR} depicts the acquired channel quality.
The \ac{SNR}, calculated using the \ac{EVM} for all measurements, is around \SI{20}{\decibel} and, thus, accurate channel predictions are possible.
The different noise figures are caused by the two different \ac{USRP} devices.

\begin{table}[H]
\centering
\caption{Results of \ac{SISO} channel prediction for measured data}
\begin{tabular}{c|c|c|c}
$\Delta f$ & Indoor \ac{LoS}      & Indoor \ac{NLoS}    & Outdoor \ac{LoS}    \\
\hline                                                                        
\multirow{ 2}{*}{\SI{25}{\mega \hertz}}     & $\delta_{h}$=0.95                           & $\delta_{h}$=0.80                             & $\delta_{h}$=0.93          \\
                                            & NMSE=\SI{-20}{\decibel}                     & NMSE=\SI{-17}{\decibel}                       & NMSE=\SI{-23}{\decibel}    \\ \hline
\multirow{ 2}{*}{\SI{50}{\mega \hertz}}     & $\delta_{h}$=0.90                           & $\delta_{h}$=0.77                             & $\delta_{h}$=0.87          \\
                                            & NMSE=\SI{-19}{\decibel}                     & NMSE=\SI{-14}{\decibel}                       & NMSE=\SI{-21}{\decibel}    \\
\end{tabular}
\label{tab:Res-Meas-SISO}
\end{table} 
\noindent
The results, in Table~\ref{tab:Res-Meas-SISO}, show that the \ac{NN} achieves reasonable precision for all scenarios studied.
As expected the indoor \ac{NLoS} scenario is worse when compared to the corresponding \ac{LoS} scenario, but, still, in a reasonable range.
One must keep in mind that these results are based on actual measurements including several impairments, e.g., caused by hardware tolerances and quantization effects. 
Thus, the used labels (i.e., measured CSI) are already compromised by noise and distortion stemming from an imperfect (but unavoidable) measurement setup.
Again, the prediction quality in all three environments turns out to be sufficiently good, suggesting deployment of the proposed system in a real world physical channel in the \ac{SISO} context.

\section{\ac{MIMO} \ac{UL}-\ac{DL} channel prediction}
\label{sec:mimo}

Next, we consider the practically more relevant \ac{MIMO} scenario, i.e., we assume multiple antennas at the \ac{BS} and try to enable \ac{FDD} Massive \ac{MIMO}, without the need of imposing any sparsity constraint on the channel.
Although multiple trained \ac{SISO} estimators can be stacked to build such a \ac{MIMO} system, we want to study whether true \ac{MIMO} signal processing (i.e., exploration of correlations between antennas) can further improve prediction accuracy.
\subsection{System Model}
We consider a single-antenna \ac{UE} and a \ac{BS} equipped with $M$ antennas.
We denote the \ac{UL} and \ac{DL} channels from the \ac{UE} to the $m$-th antenna of the \ac{BS} by $\hv^{(m)}_\text{UL}\in\mathbb C^{N_\text{sub}}$ and $\hv^{(m)}_\text{DL}\in\mathbb C^{N_\text{sub}}$, for $m=1,\dots,M$, respectively.
The investigations of Section \ref{sec:siso} are extended to the \ac{MIMO} system model in the following.
\subsection{Simulated \ac{LoS} Model}
Similar to Section \ref{sec:siso-LoS}, we first investigate the performance of a simple \ac{LoS} Massive \ac{MIMO} system.
To verify whether the channel prediction limits the gain of the linear precoding scheme, the classic \ac{MRT} is investigated.
The received signal per subcarrier $k$ is defined as
\begin{equation}
y_{\it{k}} = \hv_{\it{k}}\wv_{\it{k}}^Hs_{\it{k}} + \nv_{\it{k}}
\end{equation}
where $\wv_{\it{k}}$ is the $1\times M$ linear precoding vector, $\hv_{\it{k}}$ is the $1\times M$ channel vector, $s_{\it{k}}$ is the transmitted \ac{QPSK} symbol and $\bf{n}_{\it{k}}$ is the $1\times 1$ additive white Gaussian noise vector.
The linear precoder for \ac{MRT} is defined as 
\begin{equation}
 \wv_{\it{k}} = \frac{\hat{\hv}_{\it{k}}}{\Vert \hat{\hv}_{\it{k}}\Vert_\text{2}}.
\end{equation}

We use the same settings as in Section \ref{sec:siso} with $\Delta f=\SI{25}{\mega \hertz}$ and a \ac{NN} as depicted in Table~\ref{tab:NN-LoS-SISO} (extended to the current number of antennas $M$).
We observe a growing training complexity, yet the required amount of data points does not increase with the number of antennas.
Intuitively, this can be explained by the fact, that (per antenna) the problem itself has similar complexity as in the \ac{SISO} scenario.
However, due to the increased input dimensions, the task of finding an underlying structure 
from observations becomes more complex for larger input dimension (\emph{more randomness} in the input).

\begin{figure}[H]
	\centering
	\includegraphics{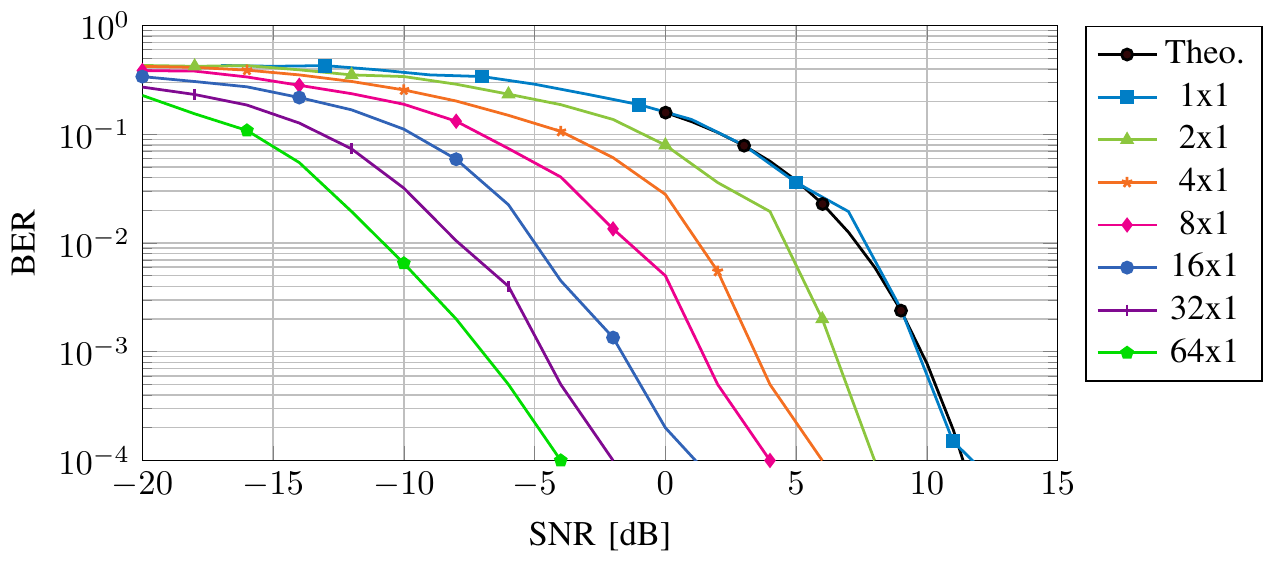}
	\caption{\ac{NN}-based \ac{CSI} prediction: \ac{BER} curves for \ac{LoS} \ac{MIMO}.}
	\label{fig:MIMO-BER-LoS}
\end{figure}

The \ac{BER} curves in Fig.~\ref{fig:MIMO-BER-LoS} show the expected \SI{3}{\decibel} gain whenever the number of antennas is doubled.
Also, the analytical \ac{BER} curve matches with the simulated 1$\times$1 curve.
Unfortunately, the joint prediction does not seem to improve performance (same \ac{NMSE} per antenna as in the \ac{SISO} scenario), indicating that the \ac{NN} does not benefit from multiple antennas (only from \ac{MRT}).
This could be explained by the fact that the antennas are assumed to be uncorrelated and, thus, each antenna needs to be predicted by itself, as the positions are also uncorrelated. 

\subsection{Results for standardized \ac{MIMO} channel models}

We now extend our \ac{NN} model as given in Table~\ref{tab:MIMO-NN-Quadrigra} to the \ac{MIMO} input.
A rectangular 8$\times$8 patch antenna array is used for all simulated scenarios (8$\times$4 for the measurement scenarios) and the \ac{UL}/\ac{DL} band separation is $\Delta f=\SI{25}{\mega \hertz}$.

\begin{table}[H]
\centering
\caption{\ac{NN} architecture for \ac{MIMO} experiments} 
\begin{tabular}{c|c}
Layers:                   & Output dimensions           \\
\hline                                                                       
Input                     & $M$ x 1024 x 2 (Re/Im)   \\
Conv1D                    & 1024 x 32                  \\
Average Pooling            & 256 x 32                  \\
Conv1D                    & 256 x16                \\
Average Pooling            & 64  x16                   \\
Flatten                   & 1024                           \\
Dense                     & 16                          \\
Dense                     & 32                       \\
Dense                     & 64                        \\
Dense                     & $M$ $\cdot$ 2048                        \\
Reshape                   & $M$ x 1024 x2  (Re/Im)   \\
\end{tabular}
\label{tab:MIMO-NN-Quadrigra}
\end{table}

Obviously, the amount of trainable parameters increases from $\sim$400,000 weights to approximately 8 million weights caused by the increased input dimensionality ($M$ antennas instead of 1).
However, somewhat to our surprise, experiments conducted with 2-D \ac{CNN} structures did not yield a significant improvement in prediction performance (occasionally even slightly worse).
One potential explanation is that, with an increased number of antennas, the environment complexity also increases, as the number of scatterers can be different.

\begin{figure}[H]
	\centering
	\includegraphics{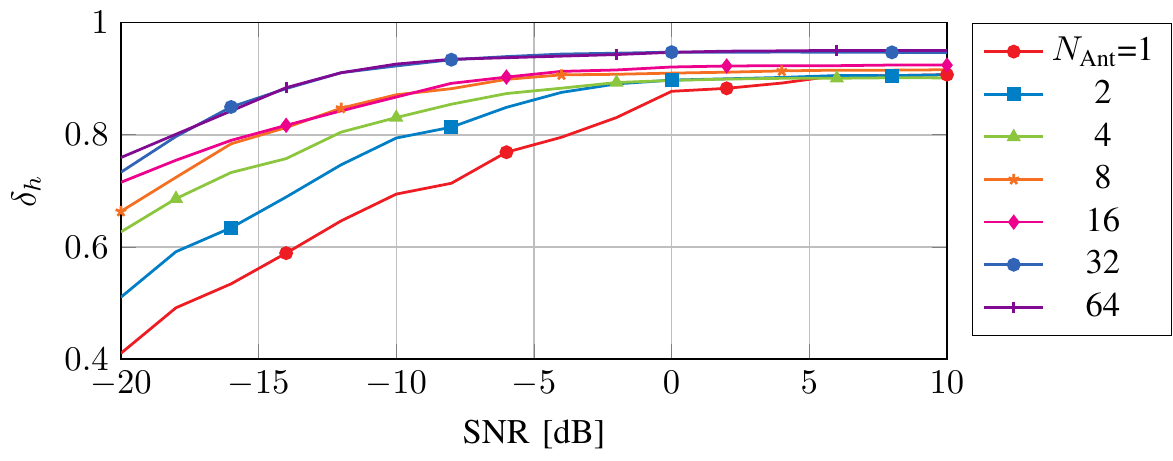}
	\caption{NN-based CSI-prediction: correlation coefficient versus SNR with different number of antennas in 3GPP \ac{LoS}.}
	\label{fig:MIMO-Corre-25MHz}
\end{figure}
The influence of different numbers of antennas on the correlation coefficient is depicted in Fig.~\ref{fig:MIMO-Corre-25MHz} for the 3GPP \ac{LoS} scenario; the training \ac{SNR} was fixed to \SI{10}{\decibel}.
It can be seen that the correlation coefficient for higher \ac{SNR} only changes slightly, but the system becomes more robust against noise with an increasing number of antennas.
Note that, due to the way we select the antennas from our dataset (without changing the aperture size), the results for 64/32 and 16/8 are similar.

\begin{figure}[H]
	\centering
	\includegraphics{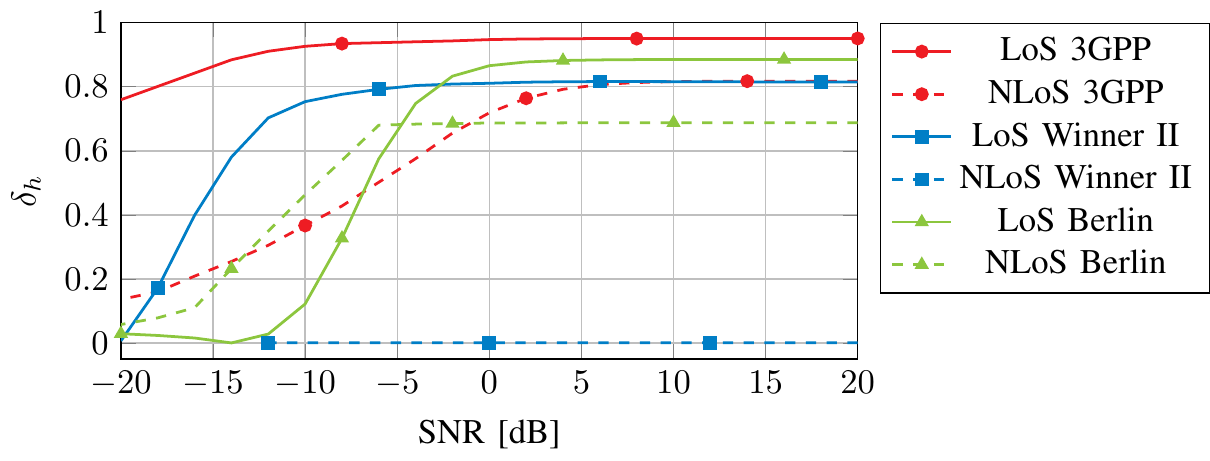}
	\caption{NN-based CSI-prediction: correlation coefficient versus \ac{SNR} with different \ac{MIMO} channels using 64 antennas.}
	\label{fig:MIMO-Models-25MHz}
\end{figure}

Fig. \ref{fig:MIMO-Models-25MHz} shows the correlation coefficient over \ac{SNR} for different Massive \ac{MIMO} channel models.
As can be seen, the prediction performance is comparable to the performance achieved in the \ac{SISO} scenario.
The prediction accuracy for the 3GPP model is within a reasonable region as the previously described margin of a correlation coefficient of $\delta_h=0.8$ is reached.
Moreover, the predictions for the \ac{NLoS} ``Berlin'' scenario show a better performance than their \ac{SISO} counterparts, as, potentially, antennas correlations could be exploited.

We note the second major result of this paper: The proposed \ac{NN} architecture can predict the \ac{DL} \ac{CSI} for the \ac{MIMO} scenario with a \ac{UL}/\ac{DL} band separation of $\Delta f=\SI{25}{\mega \hertz}$ (without requiring sparsity) and, therefore, enables \ac{FDD} Massive \ac{MIMO} for a wide range of different channels. 

\subsection{Results for actually measured \ac{MIMO} channels}
To further illustrate the scheme's robustness against hardware impairments, we use the same measurement scenarios as in Section \ref{sec:Meas-SISO}, but now for the \ac{MIMO} case.
The \emph{subband} approach, introduced in \cite{Arnold2018}, allows us to measure ``Massive'' \ac{MIMO} with 32 Antennas arranged in a 8$\times$4 patch array configuration.

\begin{figure}[H]
	\centering
	\includegraphics{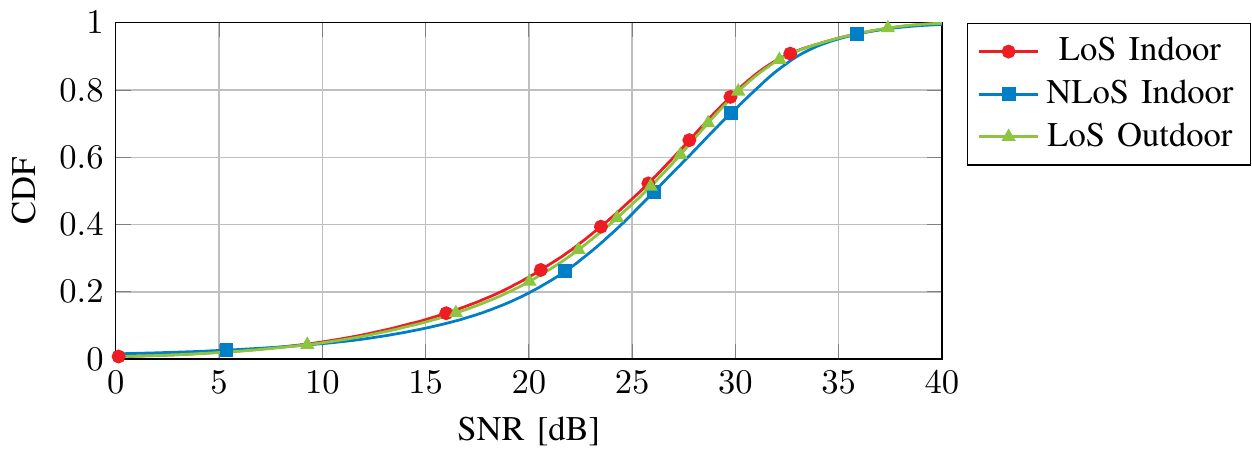}
	\caption{\ac{CDF} of the \ac{SNR} for different measurement scenarios.}
	\label{fig:MIMO-SNR-Invest}
\end{figure}

Fig. \ref{fig:MIMO-SNR-Invest} shows the \ac{CDF} of the \ac{SNR} for all antennas, and for all three measurement scenarios, respectively.
The \ac{SNR} is, on average, around \SI{25}{\decibel} resulting in useful datasets of around 5,000 spatial points per scenario. 
To evaluate the \ac{NN}'s prediction performance for this practical system we investigate the spectral efficiency by using different precoding schemes and numbers of users $N_\text{Users}$.   
The received signal signal of all \acp{UE} on the $k$th subcarrier is given by 
\begin{equation}
\yv_{k,\text{RX}}=\Hm_k \frac{\Wm_k}{\Vert \Wm_k \Vert_\text{F}} \xv_k + \nv_k
\end{equation}
where $\xv_k$ is the $N_\text{Users} \times1$ transmit vector at subcarrier $k$, $\Hm_k$ is the $N_\text{Users} \times M$ channel matrix stacked with the channel vectors $\hv_{k,1..N_\text{Users}}$, $\Wm_k$ is the $M \times N_\text{Users}$ precoding matrix, $\nv_k$ is the $N_\text{Users} \times 1$ complex gaussian noise vector, and $\yv_{\text{RX}}$ is the $N_\text{Users}\times1$ receive vector.
We precode on $N_\text{Users}$ randomly picked \ac{UE} positions of the dataset, with \acf{MRT}
\begin{equation}
\Wm_{k,\text{MR}} = \hat{\Hm}_k^{\text{H}}
\end{equation}
and, alternatively, \ac{ZF} precoding
\begin{equation}
\Wm_{k,\text{ZF}} = \hat{\Hm}_k^H \left(\hat{\Hm}_k \hat{\Hm}_k^H \right)^{-1}
\end{equation}
respectively.
With this definition, the \ac{SINR} per user $u$ can be computed as
\begin{equation}
\text{SINR}_k(u)=\frac{\left\Vert \hv_{k,u} \wv_{k,u}\right\Vert^{2}_\text{2}}{\sum_{j=1,\,j\neq u}^{N_\text{Users}}\left\Vert \hv_{k,u} \wv_{k,j}\right\Vert^{2}_\text{2}+\sigma^2}.
\end{equation}
where $\text{SINR}_{k}(u)$ is the \ac{SINR} per user $u$. 
The effective sum-rate is then calculated as
\begin{equation} 
R =\frac{1}{N_\text{sub}}\sum_{k=0}^{N_\text{sub}-1}\sum_{u=1}^{N_\text{Users}} \log_{2}\left(1+{\text{SINR}}_{k}(u)\right).
\end{equation}
\begin{figure}[H]
	\centering
	\includegraphics{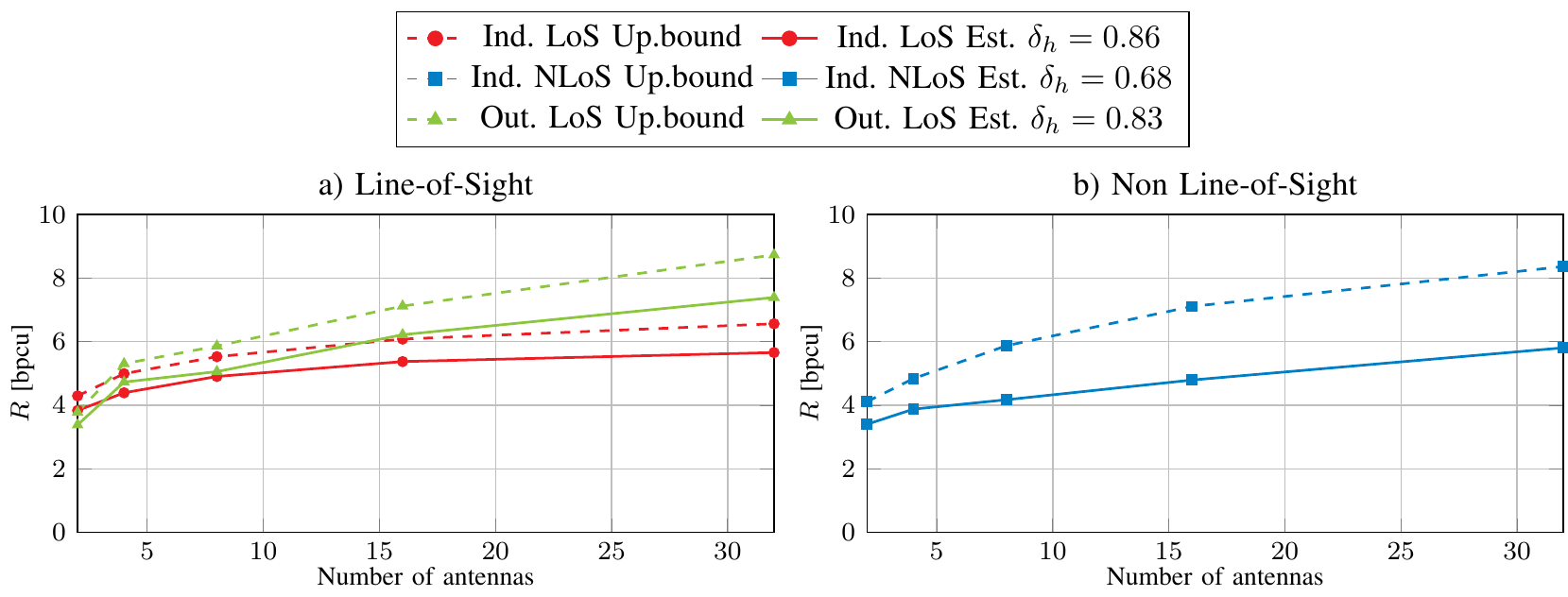}
	\caption{Simulated sum-rate versus number of antennas for the measured \ac{LoS} and \ac{NLoS} scenarios in the case of \ac{MRT} on two users; no \ac{DL} pilots used.}
	\label{fig:MIMO-Meas-Sumrate}
\end{figure}

The effective sum-rate based on the \ac{NN}'s predictions is given for all three scenarios in Fig.~\ref{fig:MIMO-Meas-Sumrate}.
We precode on two users with \ac{MRT} and try to separate them.
Note that the presented upper bound is based on the measured \ac{CSI} of the \ac{DL} frequency band, i.e., includes all hardware impairments plus noise (\ac{SNR}=\SI{10}{\decibel}).
It is, thus, not achievable with a genuine \ac{TDD} system, even if we assume perfectly reciprocal hardware.
Observe that the sum-rate based on our \ac{NN}'s predictions is only slightly worse ($\leq$10\%) than the TDD-bound in both \ac{LoS} scenarios, and is still acceptable ($\leq$20\%) in the \ac{NLoS} scenario,
reaffirming that the \ac{NN}-based system is a viable solution to enable \ac{FDD} Massive \ac{MIMO}.
Furthermore it is a notable advantage that non-reciprocal hardware chains, like the one we used for these measurements, will directly be learned by our proposed \ac{NN}-based prediction system.
This means that reasonable spectral efficiency can be reached without perfectly reciprocal hardware.
\begin{figure}[H]
	\centering
	\includegraphics{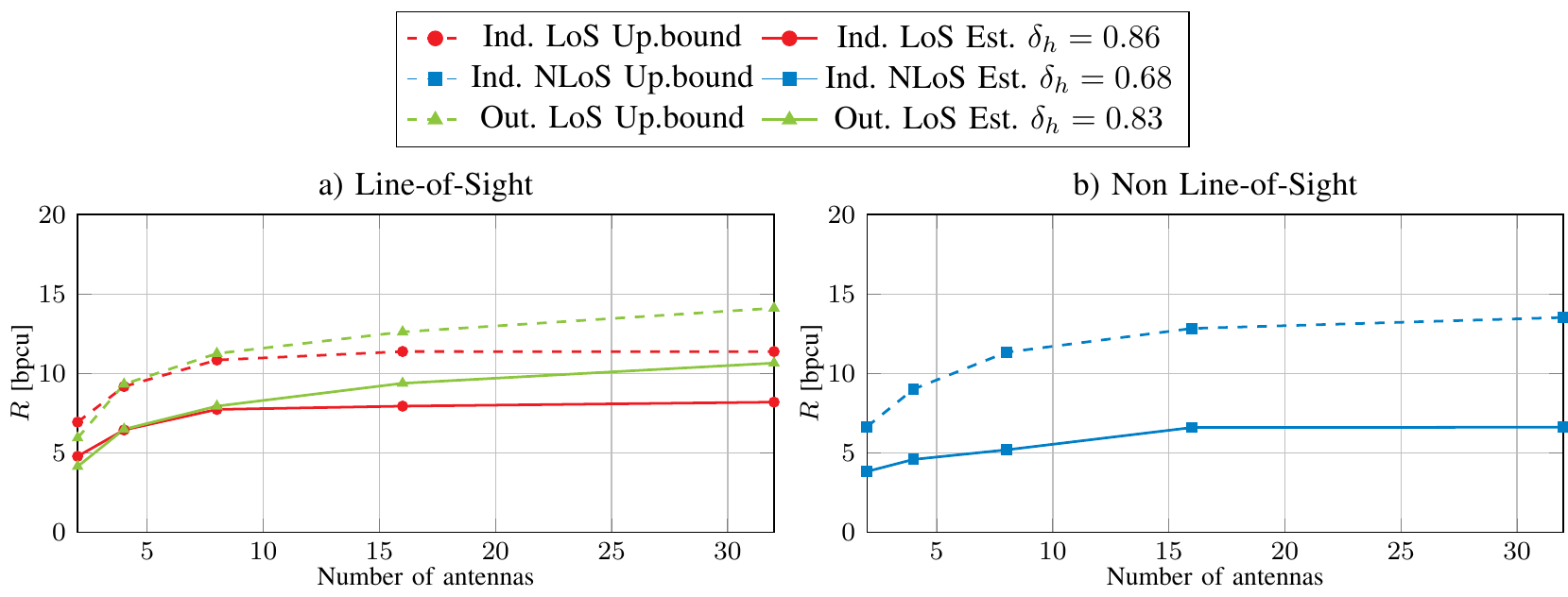}
	\caption{Simulated sum-rate versus number of antennas for the measured \ac{LoS} and \ac{NLoS} scenarios in the case of \ac{ZF} precoding on two users; no \ac{DL} pilots used.}
	\label{fig:MIMO-Meas-Sumrate_ZF}
\end{figure}

Fig.~\ref{fig:MIMO-Meas-Sumrate_ZF} provides results for the stricter case of \ac{ZF} precoding on two users with an \ac{SNR} of \SI{10}{\decibel}.
Since \ac{ZF} precoding tries to orthogonalize the users by forcing the interference to zero, a slight mismatch does have a more dramatic effect than in the simpler case of \ac{MRT}.
As can be seen, the \ac{ZF} sum-rate based on the \ac{NN}'s predictions is, again, only slightly worse ($\leq$20\%) than the TDD upper bound in both \ac{LoS} scenarios, and is still working ($\leq$45\% worse) in the \ac{NLoS} scenario.
There are several further improvements to the system at hand that may be considered in the future: a recurrent \ac{NN} could be used, an investigation into the sample density could be done, hyperparameter tuning of the \ac{NN} should be considered and a revision of the measurement setup could provide cleaner training data at even higher SNR.
Moreover, including the covariance matrix as expert knowledge could further improve the system performance.

\section{Conclusion and Outlook}\label{sec:conclusions}

We introduced a \ac{NN}-based scheme for extrapolating \ac{DL} \ac{CSI} from observed \ac{UL} \ac{CSI} in both \ac{SISO} and \ac{MIMO} scenarios, using simulated as well as actually measured channels.
The new scheme outperforms the classic Wiener filter-based approach, and, even more so, allows to operate the system in cases where the Wiener filter would fail.
Quantitatively, the \ac{NN}-based channel prediction method leads to a spectral efficiency loss 
of only $\sim$15\% in a two-user \ac{MRT} setting when compared to a \ac{TDD}-based system that is (unrealistically) assumed to be \emph{perfect} in terms of hardware reciprocity and, thus, \ac{DL} channel estimation.
Obviously, it is key to have a large training dataset available and, by its very nature, additional retraining may be required whenever the radio environment changes significantly.
It remains open for future research to further improve the \ac{NN} architecture, e.g., by using recurrent \acp{NN}, by investigating the required sample density, by applying a Kalman filter to tackle time correlation, or by adding covariance matrix knowledge to the system.
To facilitate follow-up studies, all channel measurements as well as all simulated data will be made publicly available so that results can be reproduced and improved as research in this
practically highly relevant topic further progresses\footnote{\emph{link available after review}}.

\bibliographystyle{IEEEtran}
\bibliography{IEEEabrv,references}

\begin{thebibliography}{10}
\providecommand{\url}[1]{#1}
\csname url@samestyle\endcsname
\providecommand{\newblock}{\relax}
\providecommand{\bibinfo}[2]{#2}
\providecommand{\BIBentrySTDinterwordspacing}{\spaceskip=0pt\relax}
\providecommand{\BIBentryALTinterwordstretchfactor}{4}
\providecommand{\BIBentryALTinterwordspacing}{\spaceskip=\fontdimen2\font plus
\BIBentryALTinterwordstretchfactor\fontdimen3\font minus
  \fontdimen4\font\relax}
\providecommand{\BIBforeignlanguage}[2]{{%
\expandafter\ifx\csname l@#1\endcsname\relax
\typeout{** WARNING: IEEEtran.bst: No hyphenation pattern has been}%
\typeout{** loaded for the language `#1'. Using the pattern for}%
\typeout{** the default language instead.}%
\else
\language=\csname l@#1\endcsname
\fi
#2}}
\providecommand{\BIBdecl}{\relax}
\BIBdecl

\bibitem{EmilBjoernson2017}
{E. Bj\"ornson, J. Hoydis, L. Sanguinetti}, \emph{{Massive MIMO Networks:
  Spectral, Energy, and Hardware Efficiency}}.\hskip 1em plus 0.5em minus
  0.4em\relax now publishers, Nov. 2017.

\bibitem{HoydisTenBrinkMMIMO}
{J. Hoydis, S. ten Brink, M. Debbah}, ``{Massive MIMO in the UL/DL of Cellular
  Networks: How Many Antennas Do We Need?}'' \emph{IEEE Journal on Selected
  Areas in Communications}, pp. 160 -- 171, Jan. 2013.

\bibitem{MMIMO5GCommMag}
{E. G. Larsson, O. Edfors, F. Tufvesson and T. L. Marzetta}, ``{Massive MIMO
  for next generation wireless systems},'' \emph{IEEE Communications Magazine},
  pp. 186 -- 195, Feb. 2014.

\bibitem{5GTechCommMag}
{F. Boccardi and R. W. Heath and A. Lozano and T. L. Marzetta and P. Popovski},
  ``{Five disruptive technology directions for 5G},'' \emph{IEEE Communications
  Magazine}, pp. 74 -- 80, Feb. 2014.

\bibitem{7881048}
{R. S. Ganesan and W. Zirwas and B. Panzner and K. I. Pedersen and K.
  Valkealahti}, ``{Integrating 3D Channel Model and Grid of Beams for 5G mMIMO
  System Level Simulations},'' in \emph{Vehicular Technology Conference}, Sept.
  2016, pp. 1--6.

\bibitem{DLChannelFDD}
{L. Miretti and L. G. Cavalcante and S. Stanczak}, ``{FDD massive MIMO channel
  spatial covariance conversion using projection methods},'' \emph{IEEE
  International Conference on Acoustics, Speech and Signal Processing}, April
  2018.

\bibitem{6310934}
{J. Nam and J. Ahn and A. Adhikary and G. Caire}, ``{Joint spatial division and
  multiplexing: Realizing massive MIMO gains with limited channel state
  information},'' in \emph{{46th Annual Conference on Information Sciences and
  Systems (CISS)}}, March 2012, pp. 1--6.

\bibitem{8094949}
{E. Bj{\"o}rnson, J. Hoydis, L. Sanguinetti}, ``{Massive MIMO Has Unlimited
  Capacity},'' \emph{IEEE Transactions on Wireless Communications}, pp. 574 --
  590, Nov. 2017.

\bibitem{7402270}
E.~Bj{\"o}rnson, E.~G. Larsson, and T.~L. Marzetta, ``{Massive MIMO: Ten Myths
  and One Critical Question},'' \emph{IEEE Communications Magazine}, pp. 114 --
  123, Feb. 2016.

\bibitem{8284057}
{Z. Gao and L. Dai and S. Han and C. I and Z. Wang and L. Hanzo},
  ``{Compressive Sensing Techniques for Next-Generation Wireless
  Communications},'' \emph{IEEE Wireless Communications}, pp. 144 -- 153, Feb.
  2018.

\bibitem{7008286}
{Z. Jiang and A. F. Molisch and G. Caire and Z. Niu}, ``{On the achievable
  rates of FDD massive MIMO systems with spatial channel correlation},'' in
  \emph{International Conference on Communications in China (ICCC)}, Jan. 2015,
  pp. 276--280.

\bibitem{Vasisht2016}
{D. Vasisht, S. Kumar, H. Rahul, D. Katabi}, ``{Eliminating Channel Feedback in
  Next-Generation Cellular Networks},'' in \emph{Special Interest Group on Data
  Communication (SIGCOMM)}, Aug. 2016, pp. 1--6.

\bibitem{Yang2018}
{W. Yang, L. Chen and Y. Liu}, ``{Super-resolution for Achieving Frequency
  Division Duplex (FDD) Channel Reciprocity},'' in \emph{{19th IEEE
  International Workshop on Signal Processing Advances in Wireless
  Communications}}, Dec. 2018.

\bibitem{Soltani2018}
{M. Soltani, A. Mirzaei, V. Pourahmadi and H. Sheikhzadeh}, ``{Deep
  Learning-Based Channel Estimation},'' \emph{{IEEE Communication Letters}},
  Oct. 2018.

\bibitem{8052521}
{H. Ye, G.Ye Li , B. Juang}, ``{Power of Deep Learning for Channel Estimation
  and Signal Detection in OFDM Systems},'' \emph{IEEE Wireless Communications
  Letters}, 2018.

\bibitem{ZhiyuanJiang2018}
{Z. Jiang, Z. He, S. Chen, A. F. Molisch}, ``{Inferring Remote Channel State
  Information: Cramer-Rao Lower Bound and Deep Learning Implementation},'' in
  \emph{Globecom}, Dec. 2018.

\bibitem{RuichenDeng2018}
{R. Deng, Z. Jiang, S. Zhou, S. Cuiy, and Z. Niu}, ``{A Two-Step Learning and
  Interpolation Method for Location-based Channel Database Construction},'' in
  \emph{Globecom}, Dec. 2018.

\bibitem{PeihaoDong2018}
{P. Dong, H. Zhang and G. Ye Li}, ``{Machine Learning Prediction based CSI
  Acquisition for FDD Massive MIMO Downlink},'' in \emph{Globecom}, Dec. 2018.

\bibitem{DLChannelCovarianceFDDMMIMO}
{M. Lorenzo and R. Cavalcante, and S. Stanczak}, ``{Downlink channel spatial
  covariance estimation in realistic FDD massive MIMO systems},'' \emph{IEEE
  Transactions on Wireless Communications}, pp. 574 -- 590, April 2018.

\bibitem{Jaeckel2014}
{S. Jaeckel, L. Raschkowski, K. B\"orner, and L. Thiele}, ``{QuaDRiGa: A 3-D
  multi-cell channel model with time evolution for enabling virtual field
  trials},'' \emph{IEEE Trans. Antennas Propag.}, pp. 3242 -- 3256, 2014.

\bibitem{tse2005fundamentals}
D.~Tse and P.~Viswanath, \emph{Fundamentals of wireless communication}.\hskip
  1em plus 0.5em minus 0.4em\relax Cambridge university press, 2005.

\bibitem{goodfellow2016}
I.~Goodfellow, Y.~Bengio, and A.~Courville, \emph{Deep Learning}.\hskip 1em
  plus 0.5em minus 0.4em\relax MIT Press, 2016.

\bibitem{gruber2017}
T.~Gruber, S.~Cammerer, J.~Hoydis, and S.~ten Brink, ``{On Deep Learning-Based
  Channel Decoding},'' in \emph{Proc. of CISS}, Jan. 2017, pp. 1--6.

\bibitem{simonyan2014very}
{K. Simonyan and A. Zisserman}, ``{Very deep convolutional networks for
  large-scale image recognition},'' \emph{arXiv preprint arXiv:1409.1556},
  Sept. 2014.

\bibitem{780524}
G.~Y. Li, ``{Pilot-symbol-aided channel estimation for OFDM in wireless
  systems},'' in \emph{1999 IEEE 49th Vehicular Technology Conference}, May
  1999.

\bibitem{cybenko1989approximation}
G.~Cybenko, ``{Approximation by superpositions of a sigmoidal function},''
  \emph{Mathematics of control, signals and systems}, Feb. 1989.

\bibitem{USRP}
\BIBentryALTinterwordspacing
{USRP User Manual, Ettus Research}. [Online]. Available:
  \url{https://www.ettus.com/}
\BIBentrySTDinterwordspacing

\bibitem{Arnold2018}
{M. Arnold, J. Hoydis and S. ten Brink}, ``{Novel Massive MIMO Channel Sounding
  Data Applied to Deep Learning-based Indoor Positioning},'' \emph{SCC 2019},
  Feb.

\end{thebibliography}
\end{document}